\documentclass[useAMS,usenatbib]{mn2e}
\include{epsf}

\def\gsim{\lower.73ex\hbox{$\sim$}\llap{\raise.4ex\hbox{$>$}}$\,$}
\def\lsim{\lower.73ex\hbox{$\sim$}\llap{\raise.4ex\hbox{$<$}}$\,$}
\def\etal{{et al}.}
\def\kms{km s$^{-1}$}
\def\Mpch{\, h^{-1}{\rm Mpc}}

\def\kms{\, {{\rm km}}\,{{\rm s}}^{-1} }
\def\be{\begin{equation}}
\def\ee{\end{equation}}

\title[$z$-Space Distortions in the 2QZ Survey]
  {The 2dF QSO Redshift Survey - XV. Correlation Analysis of Redshift-Space Distortions}
\author[J. da \^{A}ngela et~al.]
  {J. da \^{A}ngela$^{1,2}$, 
  P. J. Outram$^1$, T. Shanks$^1$, B.~J. Boyle$^{3,4}$, S. M. Croom$^{4}$
\newauthor   N.~S. Loaring$^{5}$, L. Miller$^{6}$ \& R.~J. Smith$^{7}$\\
1 Department of Physics, University of Durham, Science Laboratories, South Road, Durham, DH1 3LE, United Kingdom\\
2 Centro de Astrof\'{\i}sica da Universidade do Porto, R. das Estrelas s/n, 4150-762 Porto, Portugal\\
3 Australia Telescope National Facility, PO Box 76, Epping NSW 1710, Australia\\
4 Anglo-Australian Observatory, PO Box 296, Epping, NSW 2121, Australia\\
5 Mullard Space Science Laboratory, University College London, Holmbury St. Mary, Dorking, Surrey, RH5 6NT, United Kingdom\\
6 Department of Physics, University of Oxford, Keble Road, Oxford, OX1 3RH,  United Kingdom\\
7 Liverpool John Moores University, Twelve Quays House, Egerton Wharf, Birkenhead, CH41 1LD, United Kingdom\\
}

\begin{document}

\pagerange{\pageref{firstpage}--\pageref{lastpage}} \pubyear{2004}

\maketitle

\label{firstpage}

\begin{abstract}

We analyse the redshift-space ($z$-space) distortions of QSO clustering in the 2dF QSO Redshift Survey (2QZ). To interpret the $z$-space correlation function, $\xi(\sigma,\pi)$, we require an accurate model for the QSO real-space correlation function, $\xi(r)$. Although a single power-law $\xi(r) \propto r^{-\gamma}$ model fits the projected correlation function ($w_p(\sigma)$) at small scales, it implies somewhat too shallow a slope for both $w_p(\sigma)$ and the $z$-space correlation function, $\xi(s)$, at larger scales ($\gsim 20 \Mpch$). Motivated by the form for $\xi(r)$ seen in the 2dF Galaxy Redshift Survey (2dFGRS) and in standard $\Lambda$CDM predictions, we use a double power-law model for $\xi(r)$ which gives a good fit to $\xi(s)$ and $w_p(\sigma)$. The model is parametrized by a slope of $\gamma=1.45$ for $1<r<10h^{-1}$Mpc and $\gamma=2.30$ for $10<r<40h^{-1}$Mpc. As found for 2dFGRS, the value of $\beta$ determined from the ratio of $\xi(s)/\xi(r)$ depends sensitively on the form of $\xi(r)$ assumed. With our double power-law form for $\xi(r)$, we measure $\beta(z=1.4)=0.32^{+0.09}_{-0.11}$. Assuming the same model for $\xi(r)$ we then analyse the $z$-space distortions in the 2QZ $\xi(\sigma,\pi)$ and put constraints on the values of $\Omega^0_m$ and $\beta(z=1.4)$, using an improved version of the method of Hoyle \etal\ The constraints we derive are $\Omega_{m}^{0}=0.35^{+0.19}_{-0.13}$, $\beta(z=1.4) = 0.50^{+0.13}_{-0.15}$, in agreement with our $\xi(s)/\xi(r)$ results at the $\sim 1\ \sigma$ level.

\end{abstract}

\begin{keywords}
 
surveys -  quasars, quasars: general, large-scale structure of Universe, cosmology: observations

\end{keywords}

\section{Introduction}

Understanding the formation, clustering and evolution of large-scale structure (LSS) has been one of the major quests in the history of cosmology.  One of the ways of quantifying the inhomogeneity of the Universe on the largest scales is to count the number of pairs of galaxies or quasi-stellar objects (QSOs), as a function of separation between them. The resulting correlation functions from such statistical analyses appear to show a power-law behaviour for scales $\lsim 10 \Mpch$ \citep{peebles2}.
 
Statistical tools such as correlation functions or power-spectra have been used in many cosmological studies. Unfortunately, the underlying dark matter clustering may differ from that of the luminous matter, a difference that can be quantified by the bias (Kaiser 1984). Black-hole models make a wide variety of predictions for the amplitude and form of the bias (e.g. Kauffmann \& Haehnelt 2000). Elsewhere, \citet{scottnew} have tried to constrain models for the QSO bias via studies of QSO clustering evolution. In this paper, we shall use linear dynamical models to investigate QSO clustering in ``redshift-space'', with the aim of constraining the amplitude of the QSO linear bias ($b$) and the value of the cosmological density parameter $\Omega_{m}^{0}$.

To compute the distance to the QSOs one needs to assume that their observed redshifts are only due to the expansion of the Universe. However, the observed redshifts also include contributions from the peculiar motions of the QSOs. Therefore, the estimated distances are said to be measured in {\it redshift-space}, rather than in {\it real-space}. Peculiar velocities introduce distortions in the measured clustering pattern. If a given ensemble of QSOs has, on average,  a spherically symmetric clustering pattern in real space, but the QSOs have a very large velocity dispersion, then the clustering signal measured in $z$-space will be smeared along the line-of-sight. These features are commonly referred to as ``fingers-of-God''. At large scales, gravitational instabilities can also generate peculiar motions due to a coherent infall of QSOs into the potential well of higher density regions, causing a flattening of the clustering in the $z$-direction. The presence of the these dynamical distortions, namely the large-scale flattening at large scales, has been observed in galaxy surveys such as the 2dF Galaxy Redshift Survey (2dFGRS, Hawkins et al., 2003).

Besides the dynamical distortions, geometric distortions in the
$z$-space clustering pattern  also occur if a {\it wrong} cosmology is assumed to convert the measured redshifts into comoving distances. This is because the separations estimated along and across the line-of-sight have different dependences on cosmology, and hence they will change differently if incorrect cosmologies are assumed. The use of geometric distortions to study cosmology was pioneered by \citet{ap}, who  demonstrated that they are potentially a powerful cosmological test for a non-zero $\Lambda$. However, \citet{bph} noted that the geometric and dynamical effects may be degenerate.

To break this degenerate constraint between  $\Lambda$ and bias,
complementary constraints on these parameters can be considered simultaneously.
 As shown by  \citet{fiona02}, or \citet{phil04}, orthogonal constraints can be obtained from linear evolution theory of cosmological density perturbations. In this paper we develop a similar analysis using the full catalogue of the 2dF QSO Redshift Survey (hereafter 2QZ). We aim to obtain constraints on the QSO bias at $z \approx 1.4$ and on the value of the matter density of the Universe from a detailed analysis of the $z$-space distortions in the 2QZ clustering, combined with bias constraints from linear theory.

We find that, to obtain an adequate model describing the clustering along and across the line of sight, we need an accurate description of the real-space correlation function, $\xi(r)$. We develop a model for $\xi(r)$ based on the results from the 2QZ $z$-space and projected correlation functions and use it as an input to the two-dimensional $z$-space distortions model. 

The structure of this paper is as follows: in Section 2 we introduce the 2QZ sample and outline some of the most relevant aspects of the survey. In Sections 3 and 4 we start by introducing our correlation function measurements, and derive $\xi(r)$ models suitable for explaining our observations. We then compare our results with the clustering analysis of the 2dFGRS survey (Section 5) and with predictions from CDM non-linear clustering models (Section 6). In Section 7 we discuss further constraints that can be obtained on $\beta(z=1.4)$ and $\Omega_{m}^{0}$ from fitting the $z$-space distortions in the two-dimensional $z$-space correlation function. We also compare our method with that of \citet{fiona02}, in a similar study. Finally, in Section 8, we draw the conclusions of this work. 

Throughout this paper we use two different cosmologies to compute the comoving distance as a function of redshift: an Einstein-de Sitter $\Omega_{m}^{0}=1.0$ cosmology (EdS), and a flat, $\Omega_{m}^{0}=0.3$, $\Omega_{\Lambda}^{0}=0.7$ cosmology ($\Lambda$).

\section{The Data}

The 2QZ has presented the scientific community with a multitude of new and interesting results. Comprising 23338 QSOs with a limiting magnitude of $b_{J}= 20.85$, it is the largest homogeneous completed QSO survey by more than a factor of 50, at such faint magnitudes and with such high surface density (35 QSOs deg$^{-2}$) \citep{scott04}. The high number of QSOs and their spatial density make the 2QZ a powerful tool for studying QSO clustering, in order to determine the real space clustering and model the $z$-space distortions in the clustering pattern.

The survey area consists of two declination strips, $75^{\circ}\times 5 ^{\circ}$ each. One is centred on $\delta = -30^{\circ}$ and extends from $\alpha = 21^{\rm h}40^{\rm m}$ to $\alpha = 3^{\rm h}15^{\rm m}$ and the other is centred on $\delta = 0^{\circ}$ and extends from $\alpha = 9^{\rm h}50^{\rm m}$ to $\alpha = 14^{\rm h}50^{\rm m}$. (Note that these are values in the B1950 coordinate system.)

The QSOs have $18.25 < b_{\rm J} < 20.85$ and were selected using the
ultra-violet excess technique in the plane $u-b_{J}: b_{J}-r$. Due to
the colour selection used, they lie in the redshift range $0.3\lsim
z\lsim 3.0$. The number density of selected QSOs decreases abruptly
close to both redshift limits. At low redshifts, two effects contribute
to this decrease: 1) the host galaxies' colours contaminate
significantly the observed QSO colours, since faint QSOs are being
detected; 2) the host galaxy appears extended and these objects will be
missed, since only point-sources are considered in the survey. The high
redshift limit is due to the Lyman forest that, at those redshifts,
lies in the optical and therefore causes a significant absorption in
the $u$ band. The analysis performed here only includes QSOs in the
range $0.3< z< 2.2$, where the photometric completeness of the survey is $\gsim 90 \%$ \citep{scott04}. 

Around $47 \% $ of the colour - selected candidates were spectroscopically identified as QSOs, with the criterion  being the presence of broad ($> 1000 \kms$) emission lines. The remaining objects are either stars, white dwarfs, narrow-line galaxies, cataclysmic variables \citep{marsh}, or their identification was not possible from the spectra.

The spectra were obtained using the 2-degree field instrument (2dF) at the Anglo-Australian Telescope (AAT). The design of this multi-fibre spectrograph allows the acquisition of 400 spectra simultaneously over a $2^{\circ}$ diameter circular field of view \citep{lewis}. To cover the survey area with the 2dF field, a tiling algorithm that maximizes the number of galaxies and QSOs per pointing was used. The tiling of the 2dF fields has a significant effect on the angular selection function of the survey. A very detailed analysis of the angular and radial QSO selection functions used in this work can be found in \citet{phil03} and \citet{scott04}. Following \citet{phil03}, we introduce the redshift cuts $0.3<z<2.2$, which allow us to consider clustering statistics without the need for a weighting scheme. As a consequence, the final QSO catalogue used in this study comprises 19549 QSOs (8704 and 10845 QSOs in each of the strips).

\section{The redshift-space two-point correlation function}

\subsection{Estimation of The Correlation Function} 

Statistics such as the two-point correlation-function ($\xi$) can be used to describe the clustering and the distribution of QSOs. $\xi(x)$ is the excess probability (due to clustering) of finding a pair of objects at a given separation -- $x$ -- relative to a Poisson (unclustered) distribution \citep{peebles}.

\begin{equation}
\delta P = n\delta V(1+\xi(x))
\label{equation:xi_def}
\end{equation}
is the probability that one has of finding a QSO, at separation $x$ from a neighbour, located in a volume $\delta V$. $n$ is the mean number density of QSOs.

To evaluate how much the QSO distribution deviates from an unclustered set, a random ensemble of points probing the same volume as the QSOs is generated. This random distribution must describe all the characteristics of the QSO ensemble, except for its clustering, and therefore it must be generated taking into account the  same selection functions as the QSO catalogue. For details see \citet{scott04}.

To evaluate $\xi$ different estimators have been developed. Here we use the one developed by \citet{hamil93}:

\begin{equation}
\xi(x) = \frac{<DD(x)><RR(x)>}{<DR(x)>^{2}}-1,
\label{equation:xi_hamil}
\end{equation}
where $<DD(x)>$ is the number of QSO-QSO (double counted) pairs separated by $x$, $<RR(x)>$ the number of random-random pairs separated by $x$ and $<DR(x)>$ the number of QSO-random pairs also separated by $x$. 

At small and intermediate separations, where the pairs in each bin are independent, the errors on the correlation function are determined using the Poisson estimate. Their computation is quite straightforward:

\begin{equation}
\Delta \xi = (1+\xi)\sqrt{\frac{2}{<DD>}}
\label{equation:error_xi_poisson}
\end{equation}
The factor $\sqrt{2}$ arises from the fact that each pair is counted twice.
At large scales, the pairs in each separation bin are not independent,
and hence Poisson errors are not applicable and underestimate the true
errors. Following \citet{sboyle}, at large scales the errors can be
computed by $\Delta \xi = (1+\xi)\sqrt{1/N_{Q}}$, where $N_{Q}$ is the
total number of QSOs used for computing $\xi$. We use this estimate at
separation bins where the total number of pairs exceeds the number of
QSOs in the survey. Alternatively, the errors could be estimated by
considering the dispersion in either mock QSO catalogues derived from $N$-body simulations, or subsamples of our QSO ensemble. However, as shown by \citet{fiona00}, the Poisson and $N_{Q}$ errors, as used here, are in good agreement with the errors obtained with these other estimators. Note that the correlation function estimates may be correlated bin-to-bin.

\subsection{The 1-D $z$-Space Correlation Function $\xi(s)$}

\begin{figure}
\begin{center}
\centerline{\epsfxsize = 9.0cm
\epsfbox{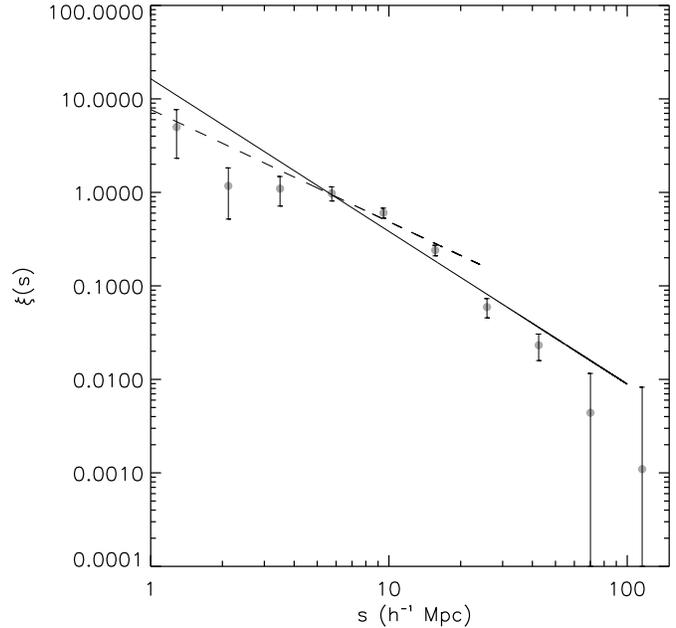}}
\caption{The $z$-space two point correlation function $\xi(s)$ for the complete 2QZ survey, measured with the Hamilton estimator, assuming $\Lambda$. The solid line represents the best fitting power-law for $1 < s < 100 \Mpch$. The dashed line is the best fitting power-law for $1 < s < 25 \Mpch$.}
\label{fig:xi_s2qz}
\end{center}
\end{figure}

A detailed description and analysis of the 2QZ $\xi(s)$ can be found in
\citet{scottnew}. They find that the 2QZ $\xi(s)$ is not well described
by a single power-law model. Fig. \ref{fig:xi_s2qz} shows the 2QZ $\xi(s)$
measured using the Hamilton estimator, assuming the $\Lambda$ cosmology
to compute the $r - z$ relation. The lines represent the best fitting
power-law models for different scales, computed by \citet{scottnew}. 
The best-fitting power-law over scales $1 < s < 100 \Mpch$\footnote{In this paper, we will denote all the separations measured in redshift space with $s$ and the ones measured in real space with $r$.}, $\xi(s) = (s/5.55)^{-1.63}$, is shown by the solid line. The dashed line represents the best fit at scales $1 < s < 25 \Mpch$; $\xi(s) = (s/5.48)^{-1.20}$. 

A noticeable feature in Fig. \ref{fig:xi_s2qz}, quantified by the discrepancy between the two best fitting power-laws, is the non-power law shape of $\xi(s)$, being flatter at small scales relative to the slope at larger separations. This small-scale flattening is frequently observed in redshift surveys and is usually attributed to the effect of $z$-space distortions, namely as a result of the QSOs' velocity dispersion in their local rest-frame. As demonstrated by \citet{scottnew}, this small scale flattening is not due to the fibre collisions of the 2dF instrument. In addition, it was also found that the majority of QSO pairs at small angular separations actually have quite large radial separations, hence not affecting the shape of $\xi(s)$ at small scales. We found that the correlation between the $\xi(s)$ bins was small. The covariance matrix shows that the correlation between adjacent bins is measured to be only $\sim 10 \%$ over the range of scales fitted (e.g. Myers et al. 2005).

\section{A Coherent Picture of the QSO clustering}

Studying the real-space clustering of the QSOs is, in itself, a subject of obvious interest: it allows the direct study of QSO clustering and its bias relative to the underlying dark matter distribution, providing a picture of how the formation and evolution of QSOs take place. Nevertheless, our motivation for studying $\xi(r)$ also includes the need for an adequate and trustworthy amplitude input for modeling the shape of the clustering in orthogonal directions. Comparing the distortions in the clustering measured along and across the line-of-sight will allow the measurement of cosmological and $z$-space distortion parameters.

Let the separations along and across the line of sight be defined as:

\be
\pi = |s_{2}-s_{1}|
\label{equation:pi_def}
\ee
\be
\sigma = \frac{(s_{1}+s_{2})}{2}\theta 
\label{equation:sig_def}
\ee
where $s_{1}$ and $s_{2}$ are the distances to two different QSOs, measured in $z$-space, and $\theta$ the angular separation between them.

Since the effects of $z$-space distortions are purely radial, the real-space clustering can be inferred from the projection of $\xi(\sigma,\pi)$ along the $\sigma$ direction\footnote{$\xi(\sigma,\pi)$ will be considered in detail in Section 7.}. The projected correlation function is obtained by integrating $\xi(\sigma,\pi)$ along the $\pi$ direction \citep{peebles}:

\be
w_{p}(\sigma) = 2\int_{0}^{\infty} \xi(\sigma,\pi) d\pi 
\label{equation:wp_def2}
\ee

When implementing this integration it must be truncated at some value $\pi_{cut}$, where $\xi$ becomes negligible. If very large scales are included, the signal amplitude will become dominated by noise. On the other hand, if the integral is only performed at very small scales, then the projected correlation function will be systematically underestimated. It was found that, in the present survey, the results did not seem sensitive to  $\pi_{cut}$ values greater than $70\ \Mpch$, which was the value used. Therefore:

\be
w_{p}(\sigma) = 2\int_{0}^{\pi_{cut}} \xi(\sigma,\pi) d\pi 
\label{equation:wp_def3}
\ee

\begin{figure}
\begin{center}
\centerline{\epsfxsize = 9.0cm
\epsfbox{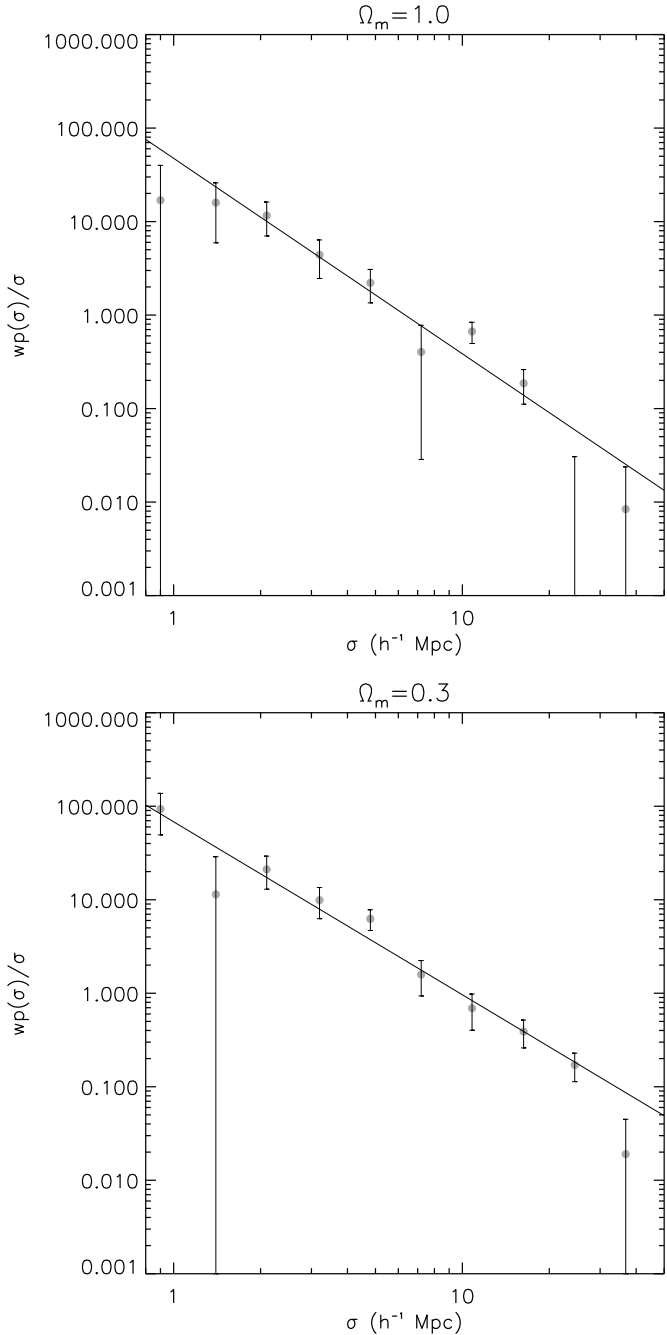}}
\caption{The projected correlation function measured with the Hamilton estimator and the best fitting power-law. The top panel shows the case where EdS was assumed, and the bottom panel the result assuming $\Lambda$. The solid lines represent the best fitting power-laws, that correspond to $\xi(r) = (r/3.77)^{-2.09}$ assuming EdS and $\xi(r) = (r/4.96)^{-1.85}$ assuming $\Lambda$.}
\label{fig:wpsig_pl}
\end{center}
\end{figure}

Fig. \ref{fig:wpsig_pl} shows $w_{p}(\sigma)/\sigma$ measured from the 2QZ catalogue. The error bars represent Poisson errors.

Since $w_{p}(\sigma)$ describes the clustering in real-space, the integral in equation \ref{equation:wp_def2} can be written in terms of the real-space correlation function $\xi(r)$ \citep{davis}:
\be
w_{p}(\sigma) = 2\int_{\sigma}^{\infty}\frac{r\xi(r)}{\sqrt{r^{2}-\sigma^{2}}}dr
\label{equation:wp_sig_xi_r}
\ee

Substituting $\xi(r)$ for a power law of the form $\left(\frac{r}{r_{0}}\right)^{-\gamma}$ in equation \ref{equation:wp_sig_xi_r}, $w_{p}(\sigma)$ will be given by:

\be
w_{p}(\sigma) =  r_{0}^{\gamma}\sigma^{1-\gamma}\left(\frac{\Gamma\left(\frac{1}{2}\right)\Gamma\left(\frac{\gamma-1}{2}\right)}{\Gamma\left(\frac{\gamma}{2}\right)}\right),
\label{equation:wp_sig_xi_r2}
\ee
where $\Gamma(x)$ is the Gamma function computed at $x$.

\begin{figure}
\begin{center}
\centerline{\epsfxsize = 9.0cm
\epsfbox{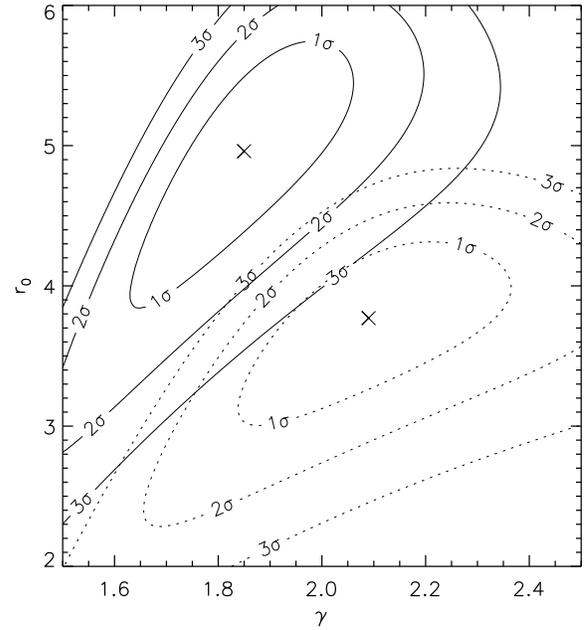}}
\caption{The best fitting values of $r_{0}$ and $\gamma$ and the respective two-parameter confidence levels obtained when fitting a model of the form  $r_{0}^{\gamma}\sigma^{1-\gamma}\left(\frac{\Gamma\left(\frac{1}{2}\right)\Gamma\left(\frac{\gamma-1}{2}\right)}{\Gamma\left(\frac{\gamma}{2}\right)}\right)$ to $w_{p}(\sigma)$. The solid contour lines represent the results from fitting the function obtained assuming $\Lambda$, and the dotted line for the EdS cosmology. The best fitting pairs of values $[r_{0},\gamma]$ are indicated with $\times$. }
\label{fig:chi2_pl_xir}
\end{center}
\end{figure}

Therefore, if the real-space correlation function $\xi(r)$ is well approximated by a power-law, then its slope will be the same as in $w_{p}(\sigma)/\sigma$, and its amplitude proportional to the $w_{p}(\sigma)/\sigma$ amplitude.
Fig. \ref{fig:wpsig_pl} also shows the best fitting power-law to
$w_{p}(\sigma)/\sigma$ (solid line). The fitting range is scaled
according to the cosmology assumed. The range used is $1 < \sigma < 25
\Mpch$, assuming EdS, and $1.2 < \sigma < 30 \Mpch$, assuming
$\Lambda$. The corresponding best fitting values of $r_{0}$ and
$\gamma$ are summarized in Table \ref{table:ro_gamma_xir}. The errors
on these parameters can be inferred from the confidence levels in the
[$r_{0}$,$\gamma$] plane, shown in Fig. \ref{fig:chi2_pl_xir}. The errors in Table \ref{table:ro_gamma_xir} and Fig. \ref{fig:chi2_pl_xir} may be underestimates because of the possibility that the $w_{p}(\sigma)$ points are correlated. However, in jacknife tests the correlation between adjacent bins is measured to be only $\sim 20 \%$ over the range of scales fitted. The correlation analysis of {\it Hubble Volume} simulations \citep{fiona00} also suggest that the $w_{p}(\sigma)$ points are approximately independent.

The close-to-unit values of the reduced $\chi^{2}$ indicate that the
power-law prediction for $\xi(r)$ corresponds to a reasonable
description of $w_{p}(\sigma)$ within the limited range it was
fitted. If EdS is assumed, then the best fitting $\xi(r)$ power-law
model has the form: $(r/3.77)^{-2.09}$, while if $\Lambda$ is assumed
instead, the best fitting power-law model is: $\xi(r) = (r/4.96)^{-1.85}$. There are deviations from the best-fitting power-law
model at scales larger than the ones used in the fit; the
$w_{p}(\sigma)$ clustering amplitude is lower than predicted  from the
best-fitting model, for $\sigma \gsim 30 \Mpch$.

\begin{table}
\centering
\begin{tabular}{||l|c|c||} \hline
                 &$\Omega_{m}^{0}=1.0$ &$ \Omega_{m}^{0}=0.3$ \\ \hline \hline
           $r_{0}\ (\Mpch)$ & $3.77^{+0.33}_{-0.47}$ &  $4.96^{+0.54}_{-0.56}$ \\\hline
           $\gamma$ & $2.09^{+0.21}_{-0.22}$ & $1.85^{+0.13}_{-0.10}$ \\\hline
           $\chi_{min}^{2}$ (reduced)& $1.61$ & $0.93$ \\\hline
           d.o.f. & $6$& $6$ \\ \hline

\end{tabular}
\caption{The values of $r_{0}$ and $\gamma$ from the best fitting
  power-law model to $w_{p}(\sigma)/\sigma$, for 6 degrees of freedom (d.o.f.), corresponding to the solid lines on Fig. \ref{fig:wpsig_pl}. If the EdS model is assumed to convert the redshifts to distances, the fit is performed between $1 \Mpch$ and $25 \Mpch$. Assuming the $\Lambda$ cosmology the range $1.2 < \sigma < 30 \Mpch$ was taken.}
\label{table:ro_gamma_xir}
\end{table}

\subsection{Does a single power-law $\xi(r)$ model explain both the $w_{p}(\sigma)$ and $\xi(s)$ results?}

The simplest explanation to account for the different slope of $\xi(s)$ at $1\lsim s \lsim 10 \Mpch$, compared to the slope at larger scales, is the effect of $z$-space distortions. The small-scale random motions of the QSOs lead to a deficit of clustering amplitude measured at scales  $\lsim 10 \Mpch$. In addition to this effect, the $\xi(s)$ clustering amplitude on large scales is affected by the infall of QSOs into the overdense regions.

According to linear theory, the relation between $\xi(s)$ and $\xi(r)$ is simply given by \citep{kaiser}:

\be
\xi(r) = \frac{\xi(s)}{1+\frac{2}{3}\beta(z)+\frac{1}{5}\beta(z)^{2}},
\label{equation:xisxir_kaiser}
\ee
where $\beta \approx \Omega_{m}^{0.6}/b$.

 Therefore, the ratio between the $z$- and real-space correlation functions gives an estimate of the infall parameter $\beta(z=1.4)$ which, for a given cosmology, allows the determination of the bias $b$.

With the current data, it is not trivial to find the real-space correlation function without making some assumptions. Inversion methods of $w_{p}(\sigma)$ like the ones described by \citet{saund} or \citet{ratcliffe} are not feasible with this survey, since the errors associated with $\xi(r)$ would be extremely large. 

\begin{figure} 
\begin{center} 
\centerline{\epsfxsize = 9.0cm
\epsfbox{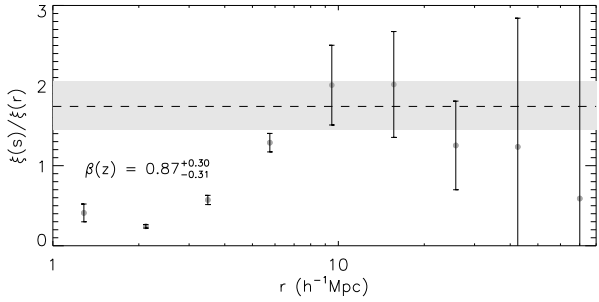}}
\caption{$\xi(s)/\xi(r)$ measured from the 2QZ survey, taking the measured $\xi(s)$ values and the single power-law $\xi(r)$ model derived, assuming $\Lambda$. See text for a full description of how the errors are computed. The fit to the function was performed on scales where the $z$-space distortions are only affected by the large-scale infall and are not contaminated by random peculiar motions of the QSOs. The dashed line and the shaded region represent this best fit and its $1\ \sigma$ confidence level. This fit corresponds to a value of $\beta(z=1.4) = 0.87^{+0.30}_{-0.31}$.}
\label{fig:xisxirbeta_1pl}
\end{center}
\end{figure}

If we approximate $\xi(r)$ by a single power-law, then the best fitting
power law $w_{p}(\sigma)$ model can be used to constrain
$\beta(z=1.4)$. This can be done by computing the ratio between the
measured values for $\xi(s)$ and the values our $\xi(r)$ model takes,
at large scales. 
Fig. \ref{fig:xisxirbeta_1pl} shows the ratio $\xi(s)/\xi(r)$ as a
function of separation.  For a given value of the separation, the
error on $\xi(s)$ can be quantified using the Poisson estimate. The
uncertainty in the $\xi(r)$ model is determined from that in the
parameters $r_0$ and $\gamma$. However, as these parameters are correlated
(see contours in Fig. \ref{fig:chi2_pl_xir}), the value found
underestimates the true uncertainty. We therefore rescale the error in $\xi(r)$ by comparison with the $\xi(s)$ Poisson errors. To estimate the error in the ratio $\xi(s)/\xi(r)$, we then add in quadrature the error in $\xi(r)$ and the Poisson errors in $\xi(s)$.
The dashed line in Fig.  \ref{fig:xisxirbeta_1pl} shows the best fitting $\xi(s)/\xi(r)$ value at scales larger then $10\ \Mpch$, and the shaded region is the $1\ \sigma$ confidence interval in the fit. The best fit produces $\beta(z=1.4) = 0.87^{+0.30}_{-0.31}$. At scales $\gsim 10 \Mpch$, non-linear effects due to peculiar velocities have a negligible effect and the $z$-space contribution comes only from the gravitational infall quantified by $\beta(z)$. This value of $\beta(z=1.4)$ is larger than other results also from the 2QZ survey, found by \citet{fiona02} and \citet{phil04}. We should point out that, using single power-law fits to the $w_{p}(\sigma)$ result of the 2dF Galaxy Redshift Survey (2dFGRS; Hawkins \etal\ 2003) also gives high values of $\beta$ from computing $\xi(s)/\xi(r)$.

\begin{figure}
\begin{center}
\centerline{\epsfxsize = 9.0cm
\epsfbox{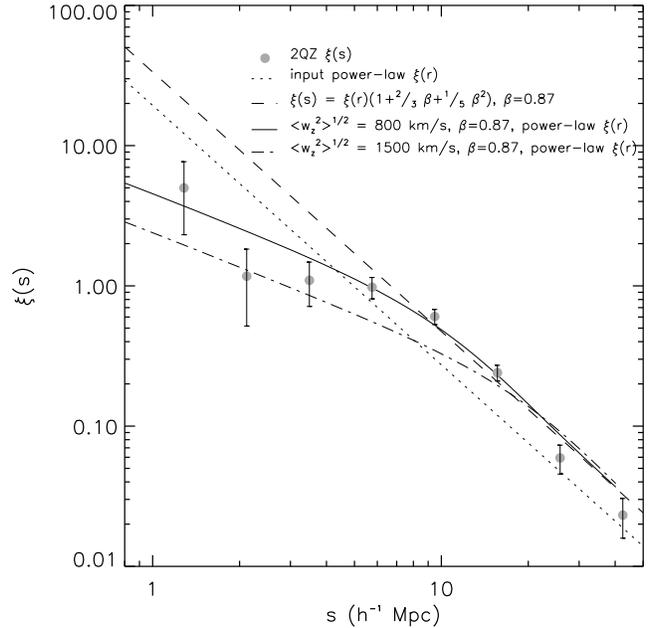}}
\caption{The circles show the measured $\xi(s)$ from the 2QZ, assuming $\Lambda$. The dotted line is the best fit single power-law $\xi(r)$ model to $w_{p}(\sigma)$. The dashed line is the same model scaled to account for the linear bias ($\beta(z)=0.87$). The solid and dash-dotted lines also include the small-scale peculiar velocities with velocity dispersion of the order $<w_{z}^{2}>^{1/2} = 800 \kms$ and  $<w_{z}^{2}>^{1/2} = 1500 \kms$, respectively.}
\label{fig:test3000}
\end{center}
\end{figure}

A model for $\xi(s)$ can then be constructed by decomposing the
separation in $\sigma$ and $\pi$ and adding the $z$-space distortions
in $\xi(\sigma,\pi)$. Then, in order to obtain $\xi(s)$,
$\xi(\sigma,\pi)$ is averaged in annuli of constant
$s=\sqrt{\sigma^{2}+\pi^{2}}$. The model used to compute
$\xi(\sigma,\pi)$ is discussed in Section
\ref{section:model1}. Fig. \ref{fig:test3000} shows the comparison of
$\xi(s)$ models with the 2QZ $\xi(s)$. The circles in the plot show the
2QZ $\xi(s)$, assuming the $\Lambda$ cosmology. The dotted line shows
the input $\xi(r)$ model, which is the best-fitting power-law to the
$w_{p}(\sigma)$. Note that this model lies well below the observed
$\xi(s)$ at all values $s \gsim 10 \Mpch$. The dashed line represents
that same power-law, scaled using equation
\ref{equation:xisxir_kaiser}, taking the value of $\beta(z) = 0.87$ to
quantify the effect of the linear $z$-space distortions. Including the distortions due to small scale pairwise velocity dispersion of the QSOs, will lead to the solid or the dash-dotted lines.
The solid line shows the effect of small-scale peculiar velocity dispersion $<w_{z}^{2}>^{1/2}=800\ \kms$. This value is dominated by the rms pairwise redshift-error of $600\ \kms$ measured from repeated observations with the remainder allowing for the intrinsic velocity dispersion of the QSOs. The effect of a larger velocity dispersion of $1500\ \kms$ is shown as the dot-dashed line.

It can be seen that the lower velocity dispersion of $800\ \kms$ gives an adequate fit to the data in the $s \lsim 10\ \Mpch$ region; the tendency of this model to overestimate the points at $2 \lsim s \lsim 4\ \Mpch$ is not resolved by moving to a bigger velocity dispersion which is seen to degrade the fit at $s \approx 10\ \Mpch $. Although the $\chi^2$ is acceptable, the slope of $\xi(s)$ at $s \gsim 10\ \Mpch$ is steeper than the single power-law model derived from $w_p(\sigma)$. It proved impossible to increase the slope in $\xi(r)$ in this range while maintaining an acceptable fit to $\xi(s)$. This tendency of single-power law $\xi(r)$ models to give poor fits to $\xi(s)$ has been seen in previous galaxy redshift surveys such as Durham/UKST  \citet{ratcliffe2} and more recently in the 2dFGRS \citet{hawk}, where a non-power-law shoulder is seen rising above the small-scale power-law at $s\approx 10\ \Mpch$, before steepening again at larger scales.

\subsection{A double power-law $\xi(r)$ model: another explanation of the $w_{p}(\sigma)$ and $\xi(s)$ results}

\citet{hawk} presented the correlation function analyses of the 2dFGRS,
both in real- and $z$-space. Including data from 166000 galaxies
with a mean redshift of  $z \approx 0.11$, this study has a statistical weight significantly higher than the 2QZ. By inverting the projected correlation function, they found that the slope of the real space correlation function of the 2dF galaxies varies with scale. In order to quantify the change in slope as a function of scale, we re-fitted their $\xi(r)$ data. At scales $\lsim 1 \Mpch$ the shape is similar to the commonly observed $\gamma \approx 1.8$ small-scales slope \citep{peebles2} however, at intermediate scales ($1 - 10 \Mpch$), that are accessible with the 2QZ survey, the $\xi(r)$ data is fitted by a shallower $\gamma = 1.45$ power-law, while at larger scales the correlation function is again steeper. Motivated by these results we now assess if the addition of a break in the $\xi(r)$ shape at $\sim 10 \Mpch$ can still provide a good description of the QSO $w_{p}(\sigma)$ and $\xi(s)$ results, whilst accounting for the issues discussed in the previous section.

\begin{figure}
\begin{center}
\centerline{\epsfxsize = 9.0cm
\epsfbox{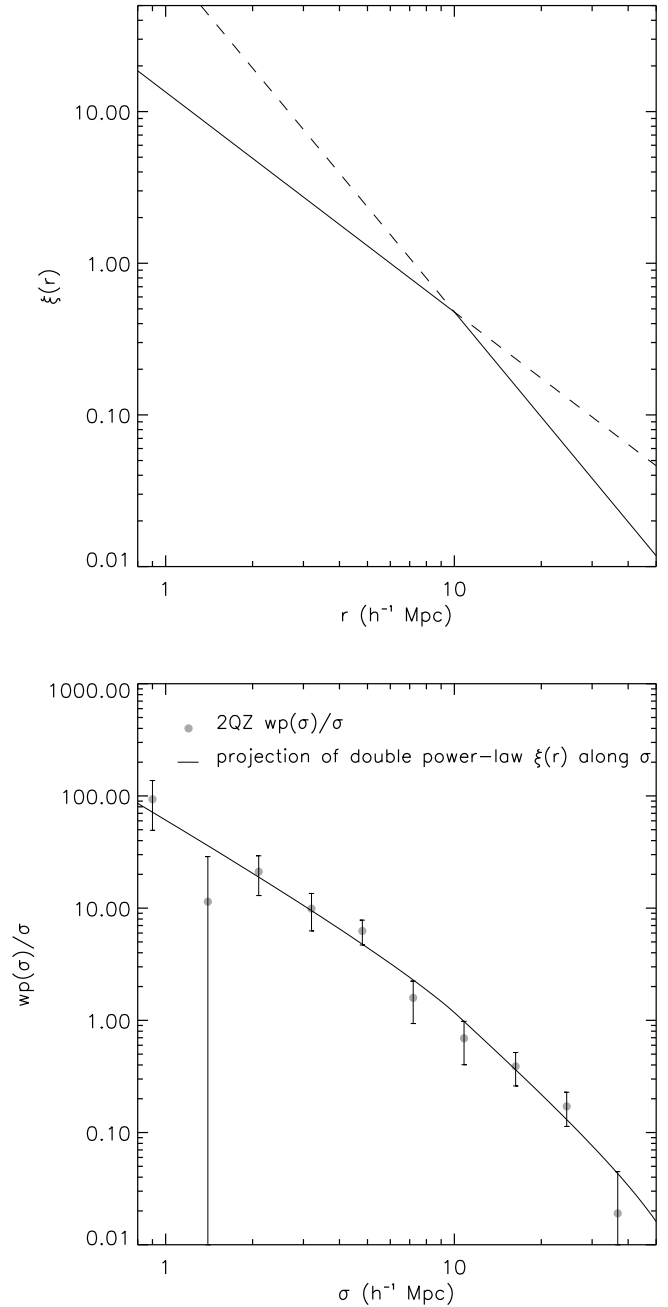}}
\caption{The top plot shows the best fitting two power-law $\xi(r)$ model to the $w_{p}(\sigma)$ data, assuming the $\Lambda$ cosmology (solid line). For scales $r < 10 \Mpch$, the power-law has the form: $\xi(r) = (r/6.0)^{-1.45}$, while for scales $r > 10 \Mpch$, it has the form $\xi(r) = (r/7.25)^{-2.30}$. The dashed lines show the extrapolation of the two power-laws to small and large scales. On the bottom, the plot shows $w_{p}(\sigma)/\sigma$ as measured from the data (circles), overplotted on the projection of the $\xi(r)$ model from the top panel (solid line).}
\label{fig:xirwpsigxis}
\end{center}
\end{figure}

We thus perform a new parametric fit to $w_{p}(\sigma)$, considering a double power-law model with a break at $10 \Mpch$ (for the $\Lambda$ cosmology). Then, for a grid of parameters that characterize the slope and amplitude of the power-laws, we project the functions describing $\xi(r)$ using equation \ref{equation:wp_sig_xi_r} to get the respective $w_{p}(\sigma)$. The best fitting model is found by performing a $\chi^{2}$ fit. The results are shown in Fig. \ref{fig:xirwpsigxis}. The value used for the upper limit of the integral in equation \ref{equation:wp_sig_xi_r} was computed from the $\pi_{cut}$ value used in equation \ref{equation:wp_def3} , to determine $w_{p}(\sigma)$ from the data.

The top plot shows the best fitting $\xi(r)$ model to $w_{p}(\sigma)$
(solid line). The parameters that describe the two power laws are: for
scales $r<10\ \Mpch$, $r_{0} = 6.0^{+0.5}_{-0.6}\ \Mpch$ and $\gamma =
1.45^{+0.27}_{-0.27}$; for $r>10\ \Mpch$, $r_{0} = 7.25\ \Mpch$ and
$\gamma = 2.30^{+0.12}_{-0.03}$. The fit is performed in the range $1.0
< \sigma < 40 \Mpch$. 
We fit the slope and amplitude of the two power-laws, keeping the break
between the two-power laws at $10 \Mpch$ and ensuring that the function
is continuous across the break. The reduced $\chi^{2}_{min}$ of this fit is $0.89$ (6 d.o.f.). The bottom plot is the measured 2QZ $w_{p}(\sigma)/\sigma$ overplotted on the projection of the best fitting $\xi(r)$ model (solid line).

This model is in very good agreement with the 2dFGRS $\xi(r)$ data, especially with the slope at intermediate scales ($1 \lsim r \lsim 10 \Mpch$). Again, we point out that with the current data we do not have enough signal at scales $\lsim 1 \Mpch$, so unfortunately we can not assess whether the 2QZ $\xi(r)$ steepens at small scales, as seen in the 2dFGRS data.

These results also show that possible deviations from a power-law $\xi(r)$ in the 2QZ are such that they do not become evident after being projected along the $\sigma$ direction. 

\begin{figure}
\begin{center}
\centerline{\epsfxsize = 9.0cm
\epsfbox{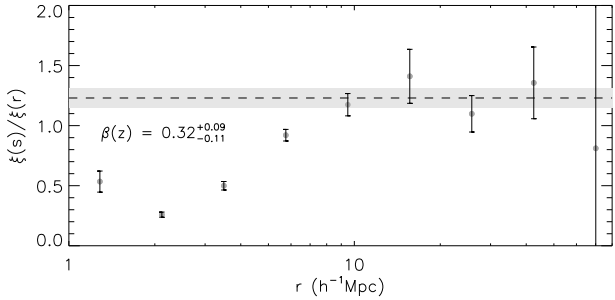}}
\caption{$\xi(s)/\xi(r)$ measured from the 2QZ survey, taking the measured $\xi(s)$ values and the two power-law $\xi(r)$ model we derived. As previously, the fit was made on scales where the $z$-space distortions are only affected by the large-scale infall and not by random peculiar motions of the QSOs. The dashed line and the shaded region represent this best fit and its $1\ \sigma$ confidence level. This fit corresponds to a value of $\beta(z=1.4) = 0.32^{+0.09}_{-0.11}$.}
\label{fig:xisxirbeta}
\end{center}
\end{figure}

The determination of $\beta$ assuming the double power-law $\xi(r)$
model can be done as described in Section 4.1. The resulting ratio
$\xi(s)/\xi(r)$ obtained as a function of separation is shown in
Fig. \ref{fig:xisxirbeta}. By fitting $\xi(s)/\xi(r)$ we find that
$\beta(z=1.4) = 0.32^{+0.09}_{-0.11}$. The quoted errors are
statistical only, and do not take into account the systematic
uncertainty in the $\beta$ measurement due to uncertainty in the form
of the $\xi(r)$ model. This is potentially quite large, as demonstrated
by the difference in the $\beta$ values obtained assuming the single
and double power-law $\xi(r)$
models.

The value of $\beta(z=1.4)$ found is in agreement with the values computed from previous estimates, with different methods and also based on QSO samples taken from the 2QZ survey (Hoyle \etal\ 2002; Outram \etal\ 2004). The value for the linear bias is $b(z=1.4) = 2.84_{-0.57}^{+1.49}$, assuming a flat $\Omega_{m}^{0}= 0.3$ cosmology. 

\begin{figure}
\begin{center}
\centerline{\epsfxsize = 9.0cm
\epsfbox{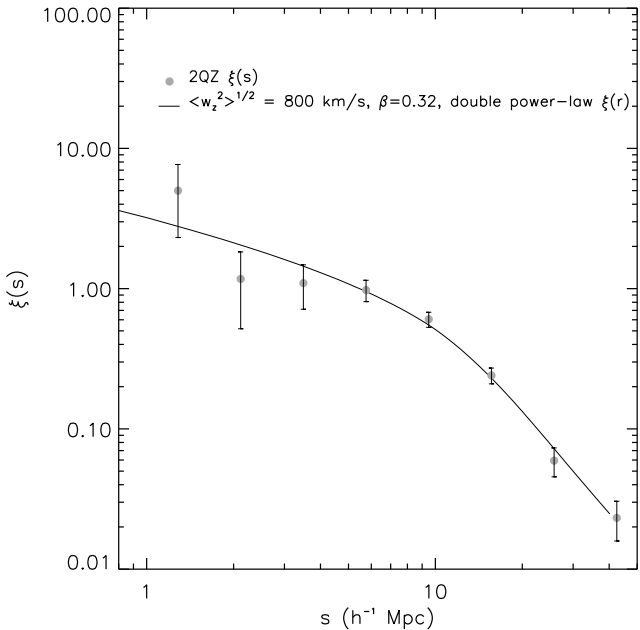}}
\caption{The plot shows the $\xi(s)$ results from the 2QZ overplotted with the $\xi(s)$ predicted by the double power-law $\xi(r)$ model, after adding the $z$-space distortions due to peculiar motions. To quantify these, we took values for $<w_{z}^{2}>^{1/2}$ and $\beta(z=1.4)$ of $800 \kms$ and $0.32$, respectively.}
\label{fig:xis_xir2pl}
\end{center}
\end{figure}

Taking the  real- and $z$-space clustering properties, the effect of $\beta$ is to change the amplitude of the correlation function at large scales. If, on top of this effect, we add the distortions due to the small-scales peculiar motions and $z$-errors, we find that the predicted $\xi(s)$ from our model is a good description of the measured $\xi(s)$ (Fig. \ref{fig:xis_xir2pl}). The reduced $\chi^{2}_{min}$ is $0.95$ (8 d.o.f.), thus indicating a very reasonable fit.

\section{Comparison with Results from Other Surveys}

The high statistical significance of the 2dFGRS sample makes it unique, and an excellent sample with which to compare our results. 
\citet{hawk} measured the $z$-space correlation function $\xi(s)$ and the projected correlation function, $w_{p}(\sigma)$. Unlike the case of the 2QZ, direct inversion of $w_{p}(\sigma)$ of the 2dFGRS allowed measuring the real-space correlation function $\xi(r)$ with great accuracy, up to scales of $\sim 20 \Mpch$.
$\xi(s)$ is very well defined by a double power-law model, up to scales $\sim 20 \Mpch$, where the data stops following the two power-law fit and rapidly tends towards zero. The flatter behaviour at small scales ($\lsim 4 \Mpch$) can be explained by the effects of peculiar velocities with dispersion $\sim 500 \kms$. 
From $\xi(s)/\xi(r)$ and $\xi(\sigma,\pi)$ fitting, a value of $\beta(z)= 0.49^{+0.09}_{-0.09}$ was found.

\begin{figure}
\begin{center}
\centerline{\epsfxsize = 9.0cm
\epsfbox{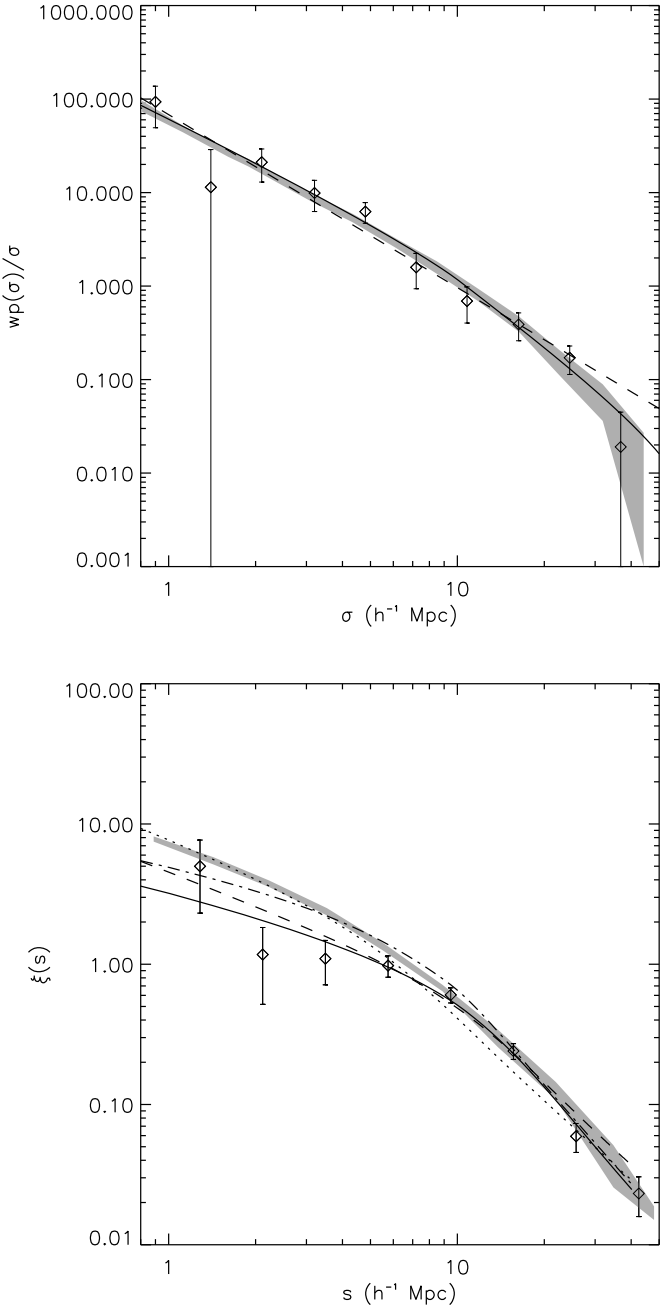}}
\caption{The top plot shows the $w_{p}(\sigma)$ results for the 2dFGRS and 2QZ surveys, together with the predictions from the single and double power-law $\xi(r)$ models previously derived for the 2QZ. The shaded region represents the $1 \sigma$ error margin for the 2dFGRS $w_{p}(\sigma)/\sigma$. The diamonds are the $w_{p}(\sigma)/\sigma$ values measured for the 2QZ. Similarly, the bottom plot shows the $\xi(s)$ results, where the shaded region is again the $1 \sigma$ error margin for the 2dFGRS results and the diamonds the 2QZ data. Refer to the text for a complete description of the models here represented.}
\label{fig:wpsigxis_2dfgrs}
\end{center}
\end{figure}

Although the double power-law model was motivated by the form of the 2dFGRS $\xi(r)$, we have proceeded by fitting the model to the 2QZ $w_{p}(\sigma)$ and $\xi(s)$. Therefore, it is now worthwhile checking how the double-power law model fitted the 2QZ data compares to the 2dFGRS correlation functions. Of course, the form of the high-$z$ QSO correlation function need not be consistent with the low-$z$ galaxy correlation function. However, assuming that the bias is scale independent and that the range of scales fitted are not affected by non-linear effects, it should be expected the form of $\xi(r)$ to be the same. We compare the measured 2dFGRS $w_{p}(\sigma)$ and $\xi(s)$ with the predictions from the $\xi(r)$ 2QZ model. This is represented in Fig. \ref{fig:wpsigxis_2dfgrs}.

The shaded region, in both plots, represents the $1 \sigma$ errors from
the results presented in \citet{hawk}. The diamonds are the measured
$w_{p}(\sigma)/\sigma$ and $\xi(s)$ from the 2QZ. The dashed and solid
lines show the predictions from the  single and double power-law
$\xi(r)$ models that best fit the 2QZ $w_{p}(\sigma)$ data. The upper
plot shows their projection along $\sigma$ and the lower plot the
respective $\xi(s)$ functions, with distortions parametrizes by
$<w_{z}^{2}>^{1/2} = 800\ \kms$ and $\beta = 0.87$ (for the dashed
line) and  $\beta = 0.32$ (for the solid line). It can be seen that the
2dFGRS $w_{p}(\sigma)$ results are very similar to the 2QZ and hence to
the projection of our $\xi(r)$ double power-law model, even at scales
($\gsim 20 \Mpch$) where the 2dFGRS data does not follow the
best-fitting single power-law model. \citet{scottnew} found that the 2dFGRS and 2QZ samples have the same clustering amplitude. Our results for $w_{p}(\sigma)$ corroborate this conclusion.

There is an offset between the 2QZ $\xi(s)$ models and the 2dFGRS data,
at small scales. This is due to the different $z$-space distortions in
the two data-sets. At large scales, a great level of consistency is
seen between the results of both surveys and the double power-law
model. The dotted and dash-dotted lines in this plot are the predicted
$\xi(s)$ functions for the 2dFGRS, taking the two $\xi(r)$ models
derived for the 2QZ (the single and double power-law $\xi(r)$ models
respectively)  and adding the distortions quantified by the values of $\beta(z=0.11)$ and $<w_{z}^{2}>^{1/2}$ found by \citet{hawk}. 

The dotted line is not a good representation of the 2dFGRS data since,
at $6 \lsim s \lsim 20 \Mpch$ scales, as it underestimates the 2dFGRS
clustering, considering the $1 \sigma$ errors. The dash-dotted line, that corresponds to the 2dFGRS $\xi(s)$ prediction assuming the double power-law model, is a better description of the clustering of the 2dF galaxies at large scales, although it underestimates the clustering at scales $\lsim 3\ \Mpch$.

\section{Comparison with CDM Model Predictions}

\citet{pdodds} presented a model to describe the evolution of the power-spectra ($P(k)$) of density fluctuations in the non-linear regime. This work was then followed by \citet{smith}, who derived a more accurate method of including non-linear effects in the power-spectrum, which can be applied to more general power spectra. This new method of modelling the power-spectrum is based on a fusion of the halo model with the scaling between linear and non-linear scales proposed by \citet{hamil91}. Smith and collaborators also presented a set of high-resolution $N$-body simulations that they used to test their mass power-spectrum formula. Their results suggest that the model for the mass power-spectrum is a good description of its clustering in a very wide range of scales, $k \sim 0.1 - 100\ h{\rm Mpc}^{-1}$.

Here we compute the real-space non-linear power-spectrum for the mass using the method described in \citet{smith}. By scaling the obtained power-spectrum with a linear bias $b$, we can obtain the predicted CDM power-spectrum for the 2QZ catalogue. We also compute the $z$-space power-spectrum by adding the $z$-space distortions to the model, as described in \citet{nelson}. After computing the real- and $z$-space power-spectrum, the respective correlation functions can be derived by a simple Fourier transform of the $P(k)$ output. 

\begin{figure}
\begin{center}
\centerline{\epsfxsize = 9.0cm
\epsfbox{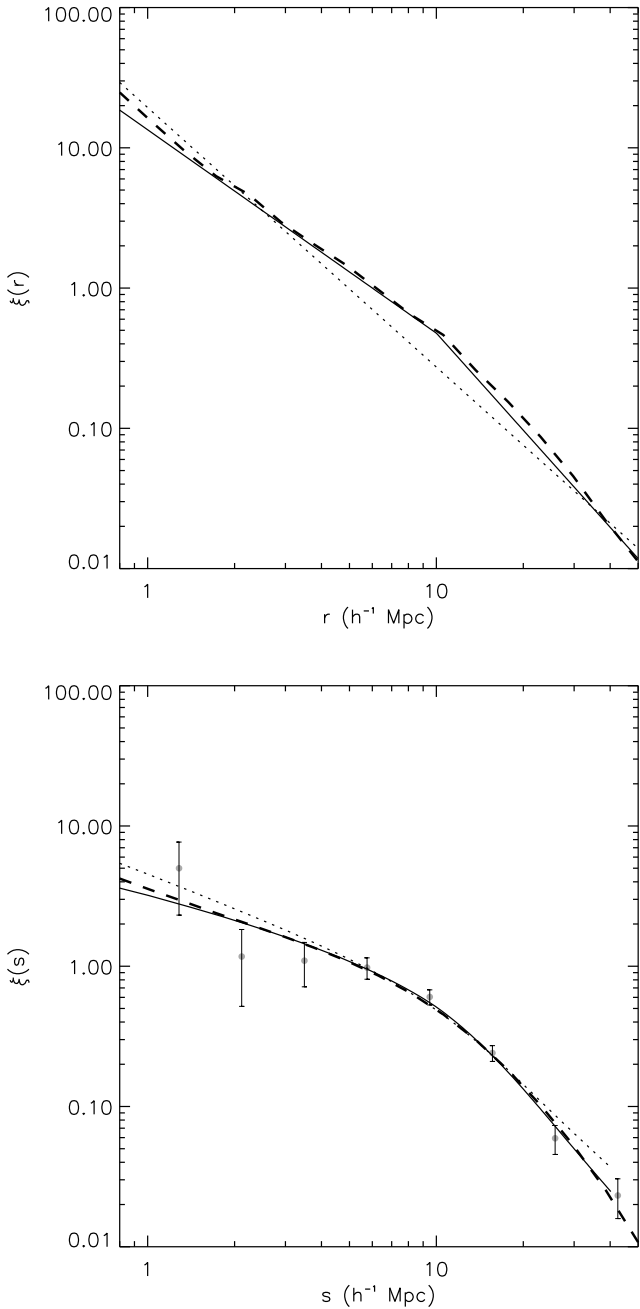}}
\caption{The top plot shows the derived biased $\xi(r)$ model from a $\Lambda$CDM $P(k)$ prediction (dashed line), overplotted on the best fitting double and single power-law $\xi(r)$ model fits to our $w_{p}(\sigma)$ data (solid line and dashed line, respectively). There is good agreement between the solid and dashed lines. In the bottom plot are the respective $\xi(s)$ functions from the models above, when the distortions parametrizes by $<w_{z}^{2}>^{1/2} = 800 \kms$ and $\beta = 0.32$ (for the double power-law model) or $\beta = 0.87$ (for the single power-law model) are added. The dashed, solid and dotted lines represent the same models as in upper plot, the difference being that now the $z$-space distortions are included. The circles are the measured $\xi(s)$ from the 2QZ data. The $\Lambda$CDM $P(k)$ predicted $\xi(s)$ shows good agreement with the data, considering the errors. Approximations made when including the distortions in the two power-law model cause a small offset between the solid and dashed lines.}
\label{fig:cdm_vs_data}
\end{center}
\end{figure}

The top plot in Fig. \ref{fig:cdm_vs_data} shows the $\xi(r)$
prediction derived from a $\Lambda$CDM power-spectrum, computed using
the method of \citet{smith} (dashed line) with parameters:
$\Omega_{m}^{0} = 0.3$, $\Omega_{\Lambda}^{0} = 0.7$, $\sigma_{8} =
0.85$, $\Gamma = 0.17$, $b= 2.3$, $<w_{z}^{2}>^{1/2} = 800 \kms$ and
$z= 1.4$. The solid line is our double power-law model that we derive
solely based on the 2QZ $\xi(s)$ and $w_{p}(\sigma)$ measurements. The
dotted line is the single power-law model derived for the 2QZ. As it
can be seen, there are reasonable similarities between the $\Lambda$CDM
prediction and the double power-law model. The slopes of the two
power-laws that describe $\xi(r)$, fitted from the $w_{p}(\sigma)$
data, are close to the predictions from $\Lambda$CDM. The single
power-law model does not appear to be as close to the
$\Lambda$CDM prediction. The bottom plot shows the $\xi(s)$ results,
after the distortions due to $\beta = 0.32$ and $<w_{z}^{2}>^{1/2} =
800 \kms$ are included. Again, the dashed line represents the
$\Lambda$CDM $P(k)$ prediction. The solid line is the $\xi(s)$ derived
from our double power-law $\xi(r)$ model, and the dotted line the
$\xi(s)$ derived from the single power-law model, once the $z$-space
distortions are added. The circles represent the $\xi(s)$ measured
directly from the 2QZ, and the corresponding error bars. The value of
$\chi^{2}_{min}$ between the $\xi(s)$ $\Lambda$CDM prediction and the
$\xi(s)$ data is $0.87$ (8 d.o.f.).  Thus, we can conclude that the
$\Lambda$CDM clustering predictions provide a good description of the
data, and that the $\Lambda$CDM real-space clustering shows more resemblance to the double power-law model than to the single power-law model for the QSO $\xi(r)$.

\section{Parameter Constraints from $z$-Space Distortions in $\xi(\sigma,\pi)$}

In Section 4 we estimated $\beta(z=1.4)$ from a direct comparison
between $\xi(s)$ and the $\xi(r)$ models derived from the
$w_{p}(\sigma)$ data. We concluded that the $\beta(z=1.4)$ value is
strongly dependent on the $\xi(r)$ model assumed. The modelling of
$z$-space distortions and their use for parameter constraints can be
taken a step further,  based on the 2-D $z$-space correlation function
$\xi(\sigma,\pi)$, developing the method of \citet{fiona02} used for the 2QZ $10k$ catalogue. 

QSO peculiar velocities lead to distortions in the $\xi(\sigma,\pi)$ shape. At small scales in $\sigma$, the random peculiar motions of the QSOs cause an elongation of the clustering signal along the $\pi$ direction. The predominant effect at large scales is the coherent infall that causes a flattening of the $\xi(\sigma, \pi)$ contours along the $\pi$ direction and some elongation along $\sigma$. 

Geometric distortions also occur if the cosmology assumed to convert the observed QSO redshifts into distances is not the same as the true, underlying cosmology of the Universe. The reason is because the cosmology dependence of the separations along the redshift direction is not the same as for the separations measured in the perpendicular direction \citep{ap}.

In the 2QZ sample it was found that, at small scales, there is not enough signal in the data to constrain the velocity dispersion of the 2dF QSOs with this method. Nevertheless, it should be possible to determine constraints not only on  $\beta(z)$, that parametrizes the large-scale infall of the QSOs, but also on the value of $\Omega_{m}^{0}$.

\subsection{The 2QZ $\xi(\sigma,\pi)$}

\begin{figure}
\begin{center}
\centerline{\epsfxsize = 9.0cm
\epsfbox{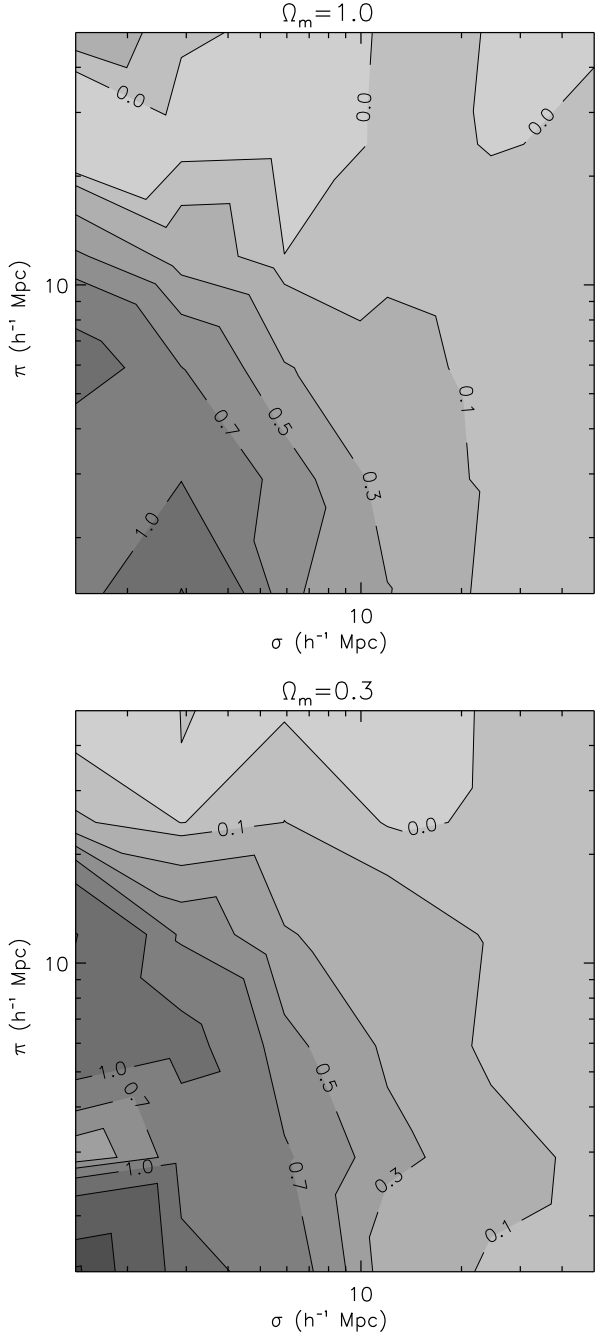}}
\caption{$\xi(\sigma,\pi)$ measured with the Hamilton estimator. On top is the result obtained if an EdS cosmology is assumed, on the bottom if the $\Lambda$ one is assumed, instead.}
\label{fig:xisigpi_2QZ}
\end{center}
\end{figure}

The measurement of $\xi(\sigma,\pi)$ is done the same way as $\xi(s)$, except that now the number of pairs is binned in two variables, rather than one. In Fig. \ref{fig:xisigpi_2QZ}  we display $\xi(\sigma,\pi)$ determined for the 2QZ catalogue. 

The $\xi(\sigma,\pi)$ contours appear elongated along $\pi$, which can
be attributed to the effects of peculiar velocities. The data is also
quite noisy, particularly at small scales in $\sigma$. The errors are
computed using the Poisson estimate. It is therefore easy to understand
the reason for the noisier contours, at those scales: since the $\sigma$
direction actually represents two dimensions, while $\pi$ only one, the
volume of each bin, and hence the number of pairs increases as the
separation $\sigma$ increases, leading to a decrease in the Poisson error.

\subsection{Comparison Between Two $z$-Space Distortions Models}

To understand and quantify the effects that are actually shaping the
$\xi(\sigma,\pi)$ contours, we develop models for the $z$-space
distortions and compare them to the observations. There are different models to explain and describe $z$-space distortions in $\xi(\sigma,\pi)$, based on different assumptions and approximations. However, one key assumption in all of them is the non-scale-dependence of the bias.

Here, two models are described and compared. The basis for the construction of the two models is the same: we start with a model for the real-space correlation function $\xi(r)$, include the effects of the coherent large-scale infall, and then convolve the result with the distribution of the small-scale pairwise peculiar velocities.

\subsection{$z$-Space Distortions - Model I}
\label{section:model1}

The large-scale coherent infall of the QSOs, is described in Fourier
space by (Kaiser 1987; Hawkins \etal\ 2003):
\be
P_{s}(k) = (1+\beta(z) \mu_{k}^{2}) P_{r}(k),
\label{equation:kaiser}
\ee
where $P_{s}(k)$ and $P_{r}(k)$ are the power-spectrum in redshift and real-space, respectively, and $\mu_{k}$ is the cosine of the angle between the wavevector $\textbf{k}$ and the line-of-sight. Translated to real-space, these results take the form (Hamilton 1992; Matsubara \& Suto 1996):

\begin{eqnarray}
\xi(\sigma,\pi) &= &\left(1+\frac{2}{3}\beta(z)+\frac{1}{5}\beta(z)^{2}\right)\xi_{0}(r)P_{0}(\mu)\\
                &- &\left(\frac{4}{3}\beta(z)+\frac{4}{7}\beta(z)^{2}\right)\xi_{2}(r)P_{2}(\mu)\\
                &+ &\frac{8}{35}\beta(z)^{2}\xi_{4}(r)P_{4}(\mu),
\label{equation:hamilton2}
\end{eqnarray}
where $\mu$ is now the cosine of the angle between $r$ and $\pi$ and $P_{l}(\mu)$ are the Legendre polynomials of order $l$. $\xi_{0}(r)$, $\xi_{2}(r)$ and $\xi_{4}(r)$ are the monopole, quadrupole and hexadecapole components of the linear $\xi(r)$ and their form will depend on the $\xi(r)$ model adopted. In general, they are given by \citep{msuto}:

\be
\xi_{2l}(r) = \frac{(-1)^{l}}{r^{2l+1}}\left(\int_{0}^{r}xdx\right)^{l}x^{2l}\left(\frac{d}{dx}\frac{1}{x}\right)^{l}x\xi(x)
\label{equation:hamilt_xigen}
\ee

The $\xi(\sigma,\pi)$ model is then convolved with the pairwise peculiar velocity distribution to include the small scale $z$-space effects due to the random motions of the QSOs. Here we assume that this can be well described by a Gaussian distribution \citep{ratcliffe}:

\be
f(w_{z}) = \frac{1}{\sqrt{2\pi}<w_{z}^2>^{1/2}}\exp\left(-\frac{1}{2}\frac{|w_{z}|^{2}}{<w_{z}^2>}\right)
\label{equation:disp_vel}
\ee

Hence, $\xi(\sigma,\pi)$ is then given by:
\be
 \xi(\sigma,\pi) = \int_{-\infty}^{\infty} \xi'(\sigma,\pi-w_{z}(1+z)/H(z)) f(w_{z}) dw_{z},
\label{equation:final_xisp}
\ee
where $\xi'(\sigma,\pi-w_{z}(1+z)/H(z))$ and $f(w_{z})$ are given by equations 12-14 and \ref{equation:disp_vel}.

\subsection{$z$-Space Distortions: Model II}
\label{section:model2}

\begin{figure*}
\begin{center}
\centerline{\epsfxsize = 16.0cm
\epsfbox{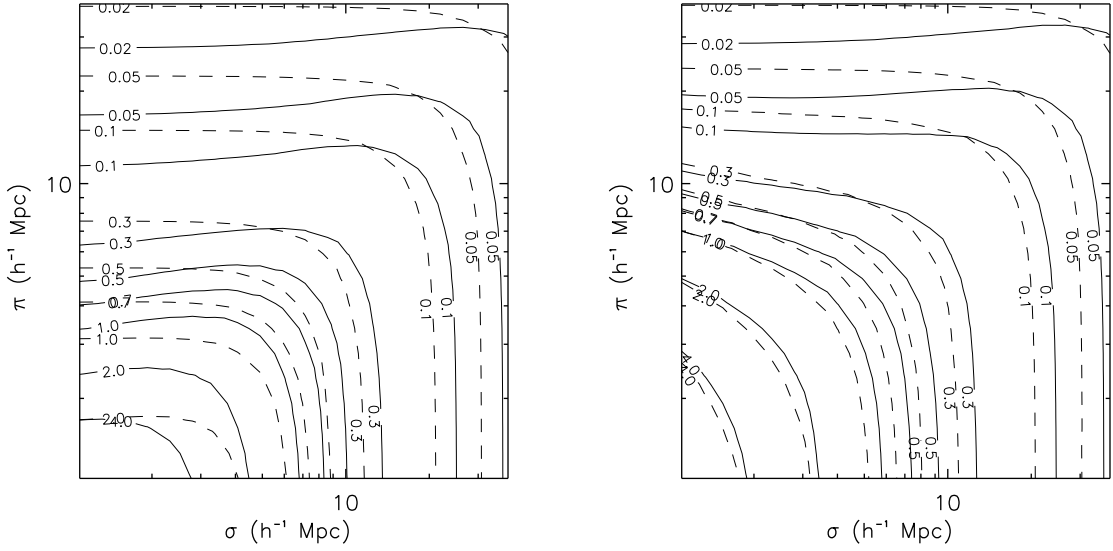}}
\caption{$\xi(\sigma,\pi)$ computed with Model I (solid line) and Model II (dashed line). The plot on the left shows the difference between the two models when  the distortions are quantified by $<w_{z}^2>^{1/2}=0 \kms$ and $\beta(z) = 0.4$. The plot on the right also accounts for the distortions caused by a non-zero velocity dispersion, and the values $<w_{z}^2>^{1/2}=800 \kms$ and $\beta(z) = 0.4$ were used. The $\xi(r)$ model is in all cases a simple power-law of the form $(r/5.0)^{-1.8}$.}
\label{fig:hoylesut_models}
\end{center}
\end{figure*}

$\xi(\sigma,\pi)$ can be defined as (Peebles 1980 ; Hoyle 2000):

\be
1+\xi(\sigma,\pi) = \int_{-\infty}^{\infty} (1+\xi(r))f(w_{z}) dw_{z}
\label{equation:hoylexisp}
\ee

Here, it is assumed that the pairwise peculiar velocity distribution $f(w_{z})$ is a slowly varying function with $r$. The form of $f(w_{z})$ is the same as in Model I and is given on equation \ref{equation:disp_vel}. The effects of the bulk motions can be included in the following way:

\be
1+\xi(\sigma,\pi) = \int_{-\infty}^{\infty} (1+\xi(r))f(w_{z}(1+z)-v(r_{z})) dw_{z},
\label{equation:hoylexisp2}
\ee
where $v(r_{z})$ is the model used for the bulk motions, as a function of the real-space separation along the $\pi$ direction  --  $r_{z}$. Following \citet{duncan} and \citet{fiona00}, 
\be
v(r_{z}) = -\frac{2}{3-\gamma}\Omega_{m}(z)^{0.6}H(z)r_{z}\frac{\xi(r)}{b^{2}+\xi(r)}
\label{equation:hoylexisp3}
\ee

$\xi(\sigma,\pi)$ computed with these two models is shown in Fig. \ref{fig:hoylesut_models}. The solid line refers to Model I and the dashed line to Model II. The input $\xi(r)$ form is the same for both models ($\xi(r) = (r/5.0)^{-1.8}$). For the plot on the right the model is computed considering  $<w_{z}^2>^{1/2}=0 \kms$ and $\beta(z) = 0.4$. The overall effect is the same for the two models, and consists of a compression of the $\xi(\sigma,\pi)$ contours along the $\pi$ direction and a small elongation along the $\sigma$ direction. The plot on the left displays the case with $<w_{z}^2>^{1/2}=800 \kms$ and $\beta(z) = 0.4$. One can see that, at small scales, the distortions caused by the velocity dispersion dominate, while at large scales, the distortions due to a non-zero $\beta(z)$ dominate. The adopted redshift for these models is $z=1.4$, which is the median redshift of the 2QZ survey. These plots also show significant differences in the shape of the distortions caused by $\beta(z)$ between the two models.

\subsection{Are Models I and II self-consistent?}
\label{section:models_xisp}

One simple test can be performed to check if a given model is self-consistent. Averaging $\xi(\sigma,\pi)$ in several annuli will give $\xi(s)$. In addition, at large scales, $z$-space distortions are mainly affected by the large-scale coherent infall, which implies the relation between $\xi(s)$ and $\xi(r)$ given by equation \ref{equation:xisxir_kaiser}. Hence, if a model is self-consistent, after averaging $\xi(\sigma,\pi)$ at constant $s$, the final result should be the same as the initial input $\xi(r)$, scaled with $(1+2/3\beta(z)+1/5\beta(z)^{2})$. If not, then the model must be wrong. Here, this test is done for Models I and II.

\begin{figure}
\begin{center}
\centerline{\epsfxsize = 9.0cm
\epsfbox{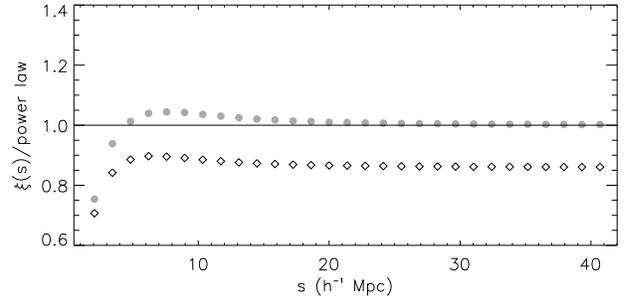}}
\caption{$\xi(s)$ computed from averaging $\xi(\sigma,\pi)$ in annuli of constant $s=\sqrt{\sigma^{2}+\pi^{2}}$ and divided by the power-law predicted at large-scales. The circles correspond to Model I and the diamonds to Model II. It was assumed $\xi(r) = (r/5.0)^{-1.8}$, $\beta(z) = 0.4$ and $<w_{z}^2>^{1/2} = 800 \kms$. $\xi(s)$ computed from Model I seems to agree very well with the prediction from the linear regime scaling from $\xi(r)$, given by equation \ref{equation:xisxir_kaiser}. The same does not happen with Model II, where an offset between the predicted and the derived $\xi(s)$ is observed. It can also be seen for both models the effect of the velocity dispersion, that  causes a flattening of $\xi(s)$ at small scales.}
\label{fig:hoylesuto_modeltest}
\end{center}
\end{figure}

Fig. \ref{fig:hoylesuto_modeltest} shows the result of this test. The circles refer to Model I and the diamonds to Model II. In both cases a $\xi(\sigma,\pi)$ model is computed using an input power-law $\xi(r) = (r/5.0)^{-1.8}$, $\beta(z) = 0.4$ and $<w_{z}^2>^{1/2}= 800 \kms$. Then, the average $\xi(\sigma,\pi)$ is computed in constant annuli of $s$, thus obtaining $\xi(s)$. The output $\xi(s)$ is then divided by the predicted $\xi(s)$ using equation \ref{equation:xisxir_kaiser}. One can see that Model I seems self-consistent, as the predicted $\xi(s)$ from $\xi(r)$ matches with the linear scaling of $\xi(r)$, through $\beta(z)$, at large scales. However, this is not the case for Model II. There is a discrepancy between the predicted  $\xi(s)$ and the one derived from averaging the $z$-space distortions on $\xi(\sigma,\pi)$. This offset corresponds to an offset on the value of $\beta(z)$ of $0.25$, which is quite significant. Given this result, only Model I will be considered, hereafter.

\subsection{Including Geometric Distortions}

It is useful to make some definitions before describing the cosmology fitting through geometric distortions. Following \citet{fiona02}, let the true, underlying cosmology of the Universe be the {\it true cosmology}, the cosmology used to build the model $\xi(\sigma,\pi)$ the {\it test cosmology}, and the cosmology assumed  to derive the $r - z$ relation, both in the model and the data to measure the correlation function, the {\it assumed cosmology}.

\begin{figure*}
\begin{center}
\centerline{\epsfxsize = 16.0cm
\epsfbox{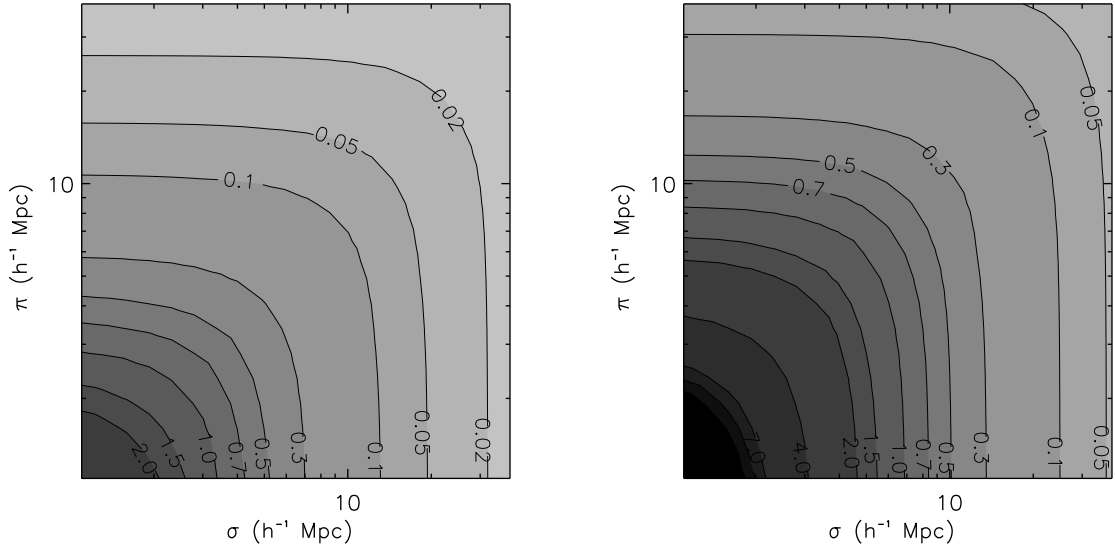}}
\caption{$\xi(\sigma,\pi)$ computed using Model I for different assumed and test cosmologies. The left plot shows the case where an EdS cosmology was assumed, to measure the correlation function, but the underlying cosmology of the Universe has $\Omega_{m}^{0} = 0.3$ and $\Omega_{\Lambda}^{0} = 0.7$. On the plot on the right is the case where the assumed cosmology is a flat cosmology with $\Omega_{m}^{0} = 0.3$ but the underlying cosmology has $\Omega_{m}^{0} = 1.0$. Assuming values for $\Omega_{m}^{0}$ higher/smaller then the true ones, will cause a flattening/elongation of the $\xi(\sigma,\pi)$ along the line-of-sight, besides the overall effect on the amplitude of the correlation function, also visible in these plots. In these cases, $z$-space distortions due to peculiar velocities were disregarded.}
\label{fig:geom_modeltest}
\end{center}
\end{figure*}

Since we are comparing the geometric distortions in both the data and the model relative to the same assumed cosmology, the test cosmology used in the model that best matches the data should be the true cosmology of the Universe, in the absence of noise.

In this paper we assume a spatially flat cosmology, and choose to fit the variable $ \Omega_{m}^{0}$, hence fixing $\Omega_{\Lambda}^{0}=1-\Omega_{m}^{0}$.

The relation between the separations $\sigma$ and $\pi$ in the test and
assumed cosmologies (referred to by the subscripts $t$ and $a$,
respectively) is the following \citep{bph}, \citep{fiona00}:

\be
\sigma_{t} = f_{\perp} \sigma_{a} = \frac{B_{t}}{B_{a}} \sigma_{a}
\label{equation:sig_cosmol}
\ee
\be
\pi_{t} = f_{\parallel} \pi_{a} = \frac{A_{t}}{A_{a}} \pi_{a}
\label{equation:pi_cosmol}
\ee
where $A$ and $B$ are defined as follows (for spatially flat cosmologies):

\be
A = \frac{c}{H_{0}}\frac{1}{\sqrt{\Omega_{\Lambda}^{0}+\Omega_{m}^{0}(1+z)^{3}}}
\label{equation:A_cosmol}
\ee

\be
B = \frac{c}{H_{0}}\int_{0}^{z}\frac{dz'}{\sqrt{\Omega_{\Lambda}^{0}+\Omega_{m}^{0}(1+z')^{3}}}.
\label{equation:B_cosmol}
\ee

In the linear regime, the correlation function in the assumed cosmology will be the same as the correlation function in the test cosmology, given that the separations are scaled appropriately. i.e.:
\be
\xi_{t}(\sigma_{t},\pi_{t}) = \xi_{a}(\sigma_{a},\pi_{a}).
\label{equation:xisp_cosmol}
\ee

In Fig. \ref{fig:geom_modeltest} the geometric distortions due to assuming a cosmology different from the true, underlying cosmology are represented. On the left plot is $\xi(\sigma,\pi)$ derived from Model I using an $\Omega_{m}^{0} = 0.3$  test cosmology, while an EdS cosmology was assumed to derive the comoving distances. The plot on the right shows the case where the underlying cosmology has $\Omega_{m}^{0} = 1.0$ but a $\Lambda$ cosmology was assumed. So, assuming a value of $\Omega_{m}^{0}$ higher than the true one will cause a compression of the $\xi(\sigma,\pi)$ contours along the line of sight, whilst  if a  too small value for $\Omega_{m}^{0}$ is assumed, it will cause an elongation of the contours along the line of sight and a high clustering amplitude to be observed. Redshift-space distortions due to the effects of peculiar velocities were not included in these models. 

Ignoring for the moment the distortions caused by the peculiar
velocities and the effects of noise, the observed shapes of the measured and modelled $\xi(\sigma,\pi)$ will be the same when the test cosmology matches the
true, underlying cosmology of the Universe (for whatever cosmology is
assumed to convert the redshifts to comoving distances). Hence, by
fitting the geometric distortions in the data, one should be able to
determine the true cosmology.

\subsection{Fitting Procedure}
\label{section:fittproc}

The fitting procedure was developed from the one used by
\citet{fiona02}. In summary, for a given value of $\beta(z)$, a
$\xi(\sigma,\pi)$ model is generated in a chosen test cosmology. Then,
the separations $\sigma$ and $\pi$ are scaled to the same cosmology
that was assumed to measure the actual data. The final model for
$\xi(\sigma,\pi)$ is then compared to the data. This method is repeated
for different test cosmologies and values of $\beta(z)$. 

Since we are fitting distortions in the {\it shape} of the $z$-space
correlation function, the correct {\it spherically-averaged amplitude}
of $\xi(r)$ must be given as an input to the model. Otherwise the fit
will be driven by offsets in the amplitude of the $\xi(\sigma,\pi)$ model from
the data, rather than the shape distortions, which would introduce
systematic errors in the constraints obtained for $\Omega_{m}^{0}$ and $\beta(z)$. 

The following steps are taken in the fitting procedure:

\begin{itemize}

\item Assume a cosmology and measure $\xi(s)$, $w_{p}(\sigma)$, $\xi(\sigma,\pi)$.

\item Take a model for the real-space correlation function, e.g. a double power-law. This model  should be a good description of the observed data, $\xi(s)$ and $w_{p}(\sigma)$.

\item Choose a pair of test values of $\Omega_{m}^{0}$ and $\beta(z)$.

\item The model for $\xi(r)$ is a good description for the data in the assumed cosmology. What is actually needed at this stage is a $\xi(\sigma,\pi)$ model in some test cosmology, hence the correct input for this model is $\xi(r)$ in that same test cosmology. Since, in the linear regime, $\xi_{t}(r_{t}) = \xi_{a}(r_{a})$, one has only to compute the real space separation in the assumed cosmology to get $\xi_{t}(r_{t})$. $r_{t}$ is given by $r_{t} = \sqrt{\sigma_{t}^2+(\pi_{t}-w_{z}/H_{t})^{2}}$ and the relation between $r_{a}$ and $r_{t}$ is: $r_{a} = r_{t}/(f_{\perp}^{2}f_{\parallel})^{1/3}$.

\item Using that model for $\xi_{t}(r_{t})$, compute $\xi_{t}(\sigma_{t},\pi_{t})$. Then, include the geometric distortions by scaling $\xi_{t}(\sigma_{t},\pi_{t})$ back to the assumed cosmology, in a similar way as described in the previous step. To get $\xi_{a}(\sigma_{a},\pi_{a})$, one needs to scale the separations $\sigma_{t}$ and $\pi_{t}$ to $\sigma_{a}$ and $\pi_{a}$, using equations \ref{equation:sig_cosmol} and \ref{equation:pi_cosmol}.

\item Adding the effects of large-scale infall not only introduces
  distortions in $\xi(\sigma,\pi)$ but also shifts the amplitude of the
  correlation function, by an amount that depends on the value of
  $\beta(z)$ taken. Since the amplitude of the spherical-averaged
  correlation function must remain the same (i.e. match the $\xi(s)$
  data), whatever $\beta$ and $\Omega_{m}^{0}$ are used as {\it test}
  values, the amplitude of the input $\xi(r)$ model is allowed to vary
  in the fit,  guaranteeing that the fit is being made to the {\it distortions} in $\xi(\sigma,\pi)$, since the averaged {\it amplitude} remains the same for whatever combination of $\beta$ and $\Omega_{m}^{0}$.

\item For the best fitting value of this amplitude factor, determine
  the $\chi^{2}$ value for the fit of this model to the data. In this step we are assuming that the $\xi(\sigma,\pi)$ bins are independent.

\item Repeat this procedure for different combinations of $\Omega_{m}^{0}$ and $\beta(z)$.

\end{itemize}

The number of degrees of freedom in the $\chi^{2}$ fit is the total
number of bins where $\xi(\sigma,\pi)$ from the model is fitted to the
data minus the number of free parameters. If the fit is to
$\Omega_{m}^{0}$ and $\beta(z)$ and the averaged amplitude of
$\xi(\sigma,\pi)$ is allowed to float so it matches $\xi(s)$, the
number of free parameters will be three. The velocity dispersion was
fixed in these fits: taking into account the $z$-errors of the survey and the intrinsic velocity dispersion of the QSOs, we assumed  $<w_{z}^2>^{1/2} = 800\ \kms$ \citep{phil03}.

\subsection{Further Constraints on $\Omega_{m}^{0}$ and $\beta(z)$ from QSO Clustering Evolution}

All the other factors being the same, the greater the value of the
true, underlying $\Omega_{m}^{0}$, the more elongated the
$\xi(\sigma,\pi)$ will be along $\pi$; and the greater the value of
$\beta(z)$, the flatter these contours will be. Hence, a degeneracy is
expected to occur in the confidence levels in the
$[\Omega_{m}^{0},\beta(z)]$ plane. In order to get better constraints
on $\Omega_{m}^{0}$ and also $\beta(z)$, this degeneracy needs to be broken. 

A possible way to break the degeneracy is to combine these results with
a constraint derived from consideration of QSO clustering evolution. From the value of the mass correlation function, at $z=0$, linear perturbation theory can be used, in a given test cosmology, to compute its value at $z=1.4$. Then, considering that the bias ($b(z=1.4)$) is given by the ratio of the QSO $\xi(r)$ and the mass $\xi(r)$, and that the former is, at large scales, related to the measured QSO $\xi(s)$ through $\beta(z=1.4)$, estimates of this parameter can be obtained for a given test cosmology.

To compute the mass correlation function amplitude at low $z$, we can use the values of $\xi(s)$ and $\beta$ found for the 2dFGRS survey. Using the values found by \citet{hawk}, the first step is to compute the respective value of $b(z=0)$.

For each test cosmology, the bias parameter of the galaxies at $z=0$ is:
\be
b(z=0) = \frac{(\Omega_{m}^{0})^{0.6}}{\beta(z=0)}
\label{equation:bias_2dfgrs}
\ee

Consider now the volume averaged two-point correlation function $\bar{\xi}^{s}$ given by:

\be
\bar{\xi}^{s} = \frac{\int_{0}^{s}4\pi s'^{2}\xi(s') ds'}{\int_{0}^{s}4\pi s'^{2} ds'}.
\label{equation:xi_bar_s}
\ee

For the 2dFGRS, $\xi(s)$ is found to be well described by a double power-law model \citep{hawk}. To compute equation \ref{equation:xi_bar_s} for the 2dFGRS, that model can be used in the integral on the numerator. Non-linear effects due to peculiar velocities in the sample should be insignificant by taking the upper limit of the integral $s = 20\ \Mpch$.
 
Then, the equivalent averaged correlation function in real-space can be determined by:
\be
\bar{\xi}^{r} (z=0) = \frac{\bar{\xi}^{s}(z=0)}{1+\frac{2}{3}\beta(z=0)+\frac{1}{5}\beta(z=0)^{2}}
\label{equation:kaiser_average}
\ee

Now the real-space mass correlation function is obtained with:

\be
\bar{\xi}^{r}_{mass} (z=0) = \frac{\bar{\xi}^{r} (z=0)}{b(z=0)^{2}},
\label{equation:bias_mass0}
\ee
where $b(z=0)$ is given by equation \ref{equation:bias_2dfgrs}.

Once determined the real-space correlation function of the mass at $z=0$, its value at $z=1.4$ is obtained using linear perturbation theory. Hence, at  $z=1.4$, the real-space correlation function of the mass will be:
\be
\bar{\xi}^{r}_{mass} (z=1.4) = \frac{\bar{\xi}^{r}_{mass} (z=0)}{G(z=1.4)^{2}},
\label{equation:bias_mass1.4}
\ee
where $G(z)$ is the growth factor of perturbations, given by linear theory and  depends on cosmology (in this case the test cosmology) \citep{cptlambda}.

Once the value of $\bar{\xi}^{r}_{mass} (z=1.4)$ is obtained for a given test cosmology, the process to find $\beta(z=1.4)$ is similar to the one used to find $\bar{\xi}^{r}_{mass} (z=0)$, but now the steps are performed backwards:

$\bar{\xi}^{s} (z=1.4)$ can be measured in a similar way as $\bar{\xi}^{s} (z=0)$. The bias factor at $z\approx 1.4$ is given by:

\be
b(z=1.4)^{2} = \frac{\bar{\xi}^{r} (z=1.4)}{\bar{\xi}^{r}_{mass} (z=1.4)},
\label{equation:bias_z1.4}
\ee
where $\bar{\xi}^{r}_{mass}$ is given by equation \ref{equation:bias_mass1.4} and $\bar{\xi}^{r} (z=1.4)$ is obtained by:

\be
\bar{\xi}^{r} (z=1.4) = \frac{\bar{\xi}^{s}(z=1.4)}{1+\frac{2}{3}\beta(z=1.4)+\frac{1}{5}\beta(z=1.4)^{2}}
\label{equation:kaiser_average2}
\ee

The value of $\beta(z=1.4)$ can then be determined by:

\be
\beta(z=1.4) = \frac{\left(\Omega_{m}(z=1.4)\right)^{0.6}}{b(z=1.4)},
\label{equation:beta_z1.4}
\ee
where $b(z=1.4)$ is given by equation \ref{equation:bias_z1.4} and $\Omega_{m}(z=1.4)$ is the value of the matter density at $z=1.4$, given by:

\be
\Omega_{m}(z) = \frac{\Omega_{m}^{0}(1+z)^{3}}{\Omega_{m}^{0}(1+z)^{3}+\Omega_{\Lambda}^{0}},
\label{equation:omegam_z1.4}
\ee
for a flat universe.

In the end, for a given value of $\Omega_{m}^{0}$ in the test cosmology, $\beta(z)$ will be obtained by solving a second order polynomial equation.

The confidence levels on the computed values of $\beta(z=1.4)$ can be obtained by considering the errors on this calculation. These are found by identifying the factors that contribute to the error, and adding the components in quadrature. Here, the components contributing to the error on $\beta(z=1.4)$ are $\beta(z=0)$, $\bar{\xi}^{s} (z=0)$ and $\bar{\xi}^{s} (z=1.4)$.

\subsection{Results}

We assume the double power-law $\xi(r)$ model to compute the
constraints on $\Omega_{m}^{0}$ and $\beta(z)$ from the
$\xi(\sigma,\pi)$ shape. By substituting this function in
Eq. \ref{equation:hamilt_xigen} we find the subsequent expressions for
the moments of the correlation function and hence the form of $\xi(\sigma,\pi)$.

It should be noted that there is some sensitivity in the $\xi(\sigma,\pi)$ fits to the detailed form of the assumed QSO $\xi(r)$. If the model for $\xi(r)$ is inaccurate then instead of fitting the distortion's {\it shape} in the $[\sigma,\pi]$ plane, the fits will be dominated by small differences in the average {\it amplitude}. As has been seen already in Section 4, the $\beta$ constraints from the $\xi(s)/\xi(r)$ ratio are also sensitive to the assumed $\xi(r)$ form. In both cases, assuming single power-law models for $\xi(r)$ would lead to higher fitted values of $\beta(z=1.4)$. We have argued that the single power-law model narrowly fails to fit the observed QSO $\xi(s)$ and $w_{p}(\sigma)$ results. As well as improving the fits to these data, the double power-law model for $\xi(r)$ is also a better representation of the correlation function results in the 2dFGRS (and Durham/UKST) galaxy survey and also in $\Lambda$CDM simulations.

\begin{figure}
\begin{center}
\centerline{\epsfxsize = 8.5cm
\epsfbox{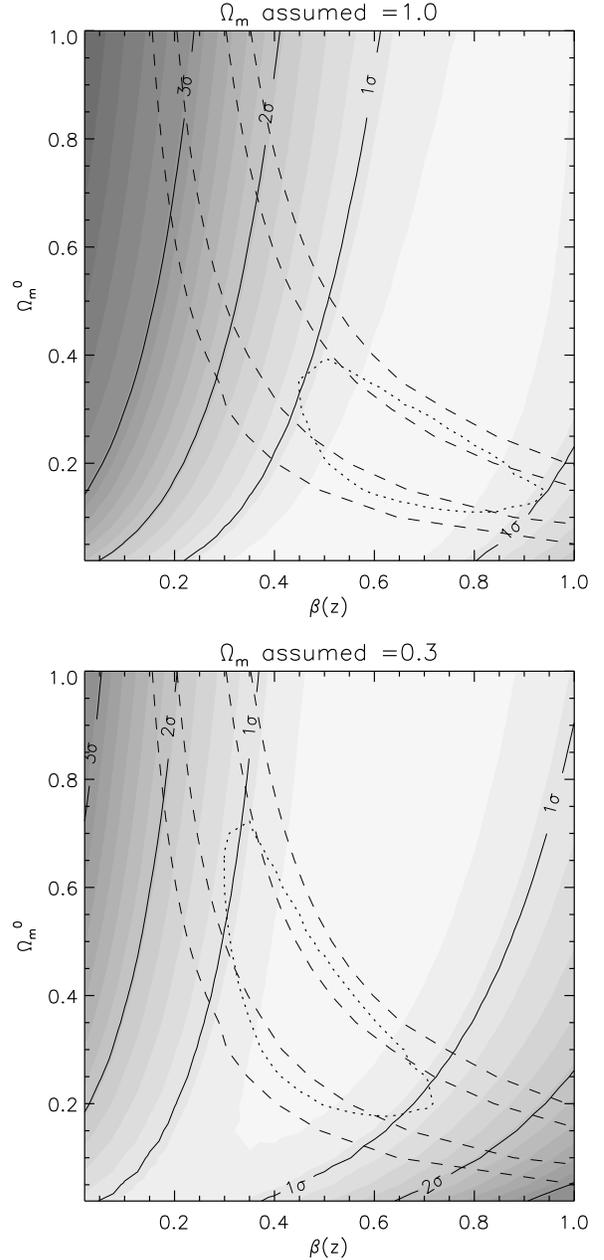}}
\caption{The confidence levels in the  $[\Omega_{m}^{0}, \beta(z)]$ plane obtained if $\xi(r)$ is described by the double power-law model, obtained from fitting the distortions in the $\xi(\sigma,\pi)$ contours (grey-scale and solid confidence levels) and from QSO clustering evolution derived from linear theory (dashed $1 \sigma$ and $2 \sigma$ confidence levels). The dotted line shows the joint two-parameter $1 \sigma$ confidence level obtained from combining both methods.}
\label{fig:xisp_beta_omega2}
\end{center}
\end{figure}

The solid lines and the shaded areas in Fig. \ref{fig:xisp_beta_omega2}
represent the results from fitting $\xi(\sigma,\pi)$ models to the
$z$-space distortions. The plots show similar likelihood contours for both the assumed cosmologies. This is as expected, since they  should in theory be independent of the assumed value of $\Omega_{m}^{0}$. The small differences that are seen are due to differences in binning the pairs between the two assumed cosmologies. It can be seen that the constraints on $\beta(z=1.4)$ are much stronger than the constraints on $\Omega_{m}^{0}$. Also, the lower the value of $\Omega_{m}^{0}$, i.e. higher $\Omega_{\Lambda}^{0}$, the more significant the effects of the geometric distortions are. This is reflected in the increased curvature of the contours for low $\Omega_{m}^{0}$. This positive slope of these contours helps breaking the degeneracy between $\Omega_{m}^{0}$ and $\beta$, when combining the results with constraints from linear growth theory (see below).

The dashed lines represent the $1 \sigma$ and $2 \sigma$ levels obtained from the bias evolution with redshift. We note that there is excellent overlap with the $1 \sigma$ constraints from $z$-distortions. Essentially, it shows that the amplitudes of mass clustering at $z \approx 1.4$, allowed by 2dFGRS dynamical analysis at $z \approx 0.1$, are consistent with the $z$-space distortions results from the QSOs, also at $z \approx 1.4$.

Finally we combine both the constraints from $z$-distortions and evolution to obtain the joint constraint represented by the dotted line. It shows the joint two-parameter $1 \sigma$ confidence level obtained with both methods.

If the EdS cosmology is assumed, the best fitting values for
$\Omega_{m}^{0}$ and $\beta(z=1.4)$ from the joint constraints are:
$\Omega_{m}^{0}=0.20^{+0.06}_{-0.11}$ and $\beta(z=1.4) =
0.70^{+0.15}_{-0.18}$, with a reduced $\chi^{2}_{min}$ of $1.10$ (12
d.o.f). If the $\Lambda$ cosmology is assumed instead, the best fitting
values for $\Omega_{m}^{0}$ and $\beta(z=1.4)$ will be:
$\Omega_{m}^{0}=0.35^{+0.19}_{-0.13}$ and $\beta(z=1.4) =
0.50^{+0.13}_{-0.15}$, with a reduced $\chi^{2}_{min}$ of $1.05$ (17
d.o.f.). Making the assumption of a double power-law model and the $\Lambda$
 cosmology, the
$\beta(z)$ constraint obtained from fitting the shape of
$\xi(\sigma,\pi)$, $\beta(z=1.4) =
0.50^{+0.13}_{-0.15}$, is slightly higher than the value of 
 $\beta(z=1.4) = 0.32^{+0.09}_{-0.11}$ obtained from fitting
 $\xi(s)/\xi(r)$, but these results are consistent within the margin of error.

Our results are also consistent with previous estimates of $\beta(z=1.4)$ and $\Omega_{m}^{0}$ from the 2QZ. This can be seen by comparing Fig. \ref{fig:xisp_beta_omega2} with Fig. 9 of \citet{fiona02}. However, the $\xi(\sigma,\pi)$ analysis presented here supersedes that of \citet{fiona02}, not only because of the $\approx 2 \times$ larger QSO data set but also because of our improved $z$-space distortion analysis. More recently \citet{phil04} used the power spectrum to study the $z$-space distortions in the 2QZ clustering. Their method probes larger scales than those in this work. Their constraints of $\beta(z=1.4)=0.45^{+0.09}_{-0.11}$ and $\Omega_{m}^{0} = 0.29^{+0.17}_{-0.09}$ are also in very good agreement with our estimate. 

\section{Conclusions}

To model $z$-space correlation functions to constrain $\Omega_{m}^{0}$ and $\beta(z=1.4)$, we have found that we need an accurate description of the real-space QSO correlation function. Although a simple power-law model gives a reasonable fit to $w_{p}(\sigma)$, we find other evidence that the real-space correlation function, $\xi(r)$, may show non-power-law features. We have argued that fitting $w_{p}(\sigma)$ and $\xi(s)$ simultaneously requires more than just a single power-law moderated by the effects of peculiar velocities. Further motivated by the non-power-law $\xi(r)$ shape measured for the 2dFGRS survey, we have found that a similar double power-law model can explain the observed 2QZ $\xi(s)$ and $w_{p}(\sigma)$. The parameters of the best-fitting model are: $r_{0}=6.0^{+0.5}_{-0.6} \Mpch$ and $\gamma = 1.45^{+0.27}_{-0.27}$, for $r<10 \Mpch$, and  $r_{0}=7.25 \Mpch$ and $\gamma = 2.30^{+0.12}_{-0.03}$, for $r>10 \Mpch$ (assuming $\Lambda$).

We then compared both the single and double power-law $\xi(r)$ models
with the 2dFGRS results and theoretical predictions from CDM clustering
models. When comparing the 2dFGRS and 2QZ projected correlation
functions, we found that our proposed double power-law model lies
within the 2dFGRS errors, which suggests that this 2QZ $\xi(r)$ model
provides an acceptable description of the 2dFGRS $\xi(r)$. With the
assumed $\Lambda$ cosmology, even the amplitudes of the 2QZ and 2dFGRS
correlation functions are similar; with the assumed EdS cosmology the
2QZ amplitude is a factor  of two smaller at $\sim 10 \Mpch$
scales. The mass correlation function amplitude evolves with redshift,
and the same is probably true for galaxies and QSOs
\citep{scottnew}. Therefore, the agreement of the amplitudes in the $\Lambda$ case is probably a coincidence. However, the 2QZ and 2dFGRS correlation functions are certainly similar in their shape.

The $\Lambda$CDM $\xi(s)$ prediction appears to be a good fit to the
2QZ $\xi(s)$. We also found that the $\xi(r)$ double power-law model
that we derived from the 2QZ clustering results is a very good match to
$\Lambda$CDM $\xi(r)$. On the other hand, the single power-law 2QZ $\xi(r)$ model, derived from the $w_{p}(\sigma)$ data,  produces a poorer match to the 2dFGRS data and the $\Lambda$CDM predictions. 

Using the double power-law  $\xi(r)$ model and the $\xi(s)$ measurements, we estimated $\beta(z=1.4)=0.32^{+0.09}_{-0.11}$, by computing and fitting $\xi(s)/\xi(r)$. This result contrasts with the higher value of $\beta(z=1.4)=0.87^{+0.30}_{-0.31}$, found when we consider the single power-law $\xi(r)$ model. Similar effects are seen in 2dFGRS, where the single power-law $\xi(r)$ model is also rejected by the data.

Modelling the $z$-space distortions in $\xi(\sigma,\pi)$ allows
constraints on $\Omega_{m}^{0}$ and $\beta(z=1.4)$ to be derived. For
the analysis we have assumed a pairwise velocity dispersion of $800
\kms$. Since this is dominated by $z$-errors ($\pm 600 \kms$ pairwise)
our results should be robust to reasonable variations in this
parameter; the pairwise velocity dispersion for the mass, for
$\Omega_{m}^{0}=0.3$ at $z = 1.4$, is $<w_{z}^{2}>^{1/2} \approx 400
\kms$ \citep{fiona00} and under simple assumptions this may only
increase to $\approx 700 \kms$, for $\Omega_{m}^{0}=1.0$.

We compared the $\xi(\sigma,\pi)$ $z$-space distortions model with that previously used by \citet{fiona02}, and found that the latter suffers from a systematic offset between the amplitude of the input $\xi(s)$ and the $\xi(\sigma,\pi)$ amplitude. This does not seem to occur in the model used in the current work \citep{msuto}.

Although the $\xi(\sigma,\pi)$ fitting is sensitive to the form of the $\xi(r)$ model, assuming the double power-law model derived above, we have obtained useful constraints on $\beta(z=1.4)$ from the $z$-space distortions analysis. The constraint on $\Omega_{m}^{0}$ is not as strong from $z$-space distortions alone, being quite degenerate with the constraints on $\beta(z=1.4)$. Combining these $z$-space distortions results with those from QSO bias evolution helps to break this degeneracy and provides stronger constraints. The resulting constraints are $\Omega_{m}^{0}=0.35^{+0.19}_{-0.13}$ and $\beta(z=1.4) = 0.50^{+0.13}_{-0.15}$, assuming the $\Lambda$ cosmology. Assuming the EdS cosmology instead produces similar results. These values are in very good agreement with the results found by \citet{phil04}. 

\section*{Acknowledgments}

The 2dF QSO Redshift Survey (2QZ) was compiled by the 2QZ survey team from observations made with the 2-degree Field on the Anglo-Australian Telescope. JA acknowledges financial support from FCT/Portugal
through project POCTI/FNU/43753/2001 and from the European Community's
Human Potential Program under contract HPRN-CT-2002-00316,
SISCO. PJO acknowledges the support of a PPARC Fellowship. We thank Catarina Lobo, Nelson Padilla and Carlton Baugh for useful comments.

\label{lastpage}

\end{document}